\documentclass[12pt,a4paper]{article}

\usepackage{epsfig}
\usepackage{color}
\usepackage{graphicx}
\usepackage{times}
\usepackage{multirow}
\usepackage{rotating}

\begin{document}

\title{State-of-the-Art Routing Protocols for Delay Tolerant Networks}
\author{Zhenxin Feng and Kwan-Wu Chin\\ {\em School of Electrical, Computer, and Telecommunications Engineering}\\ {\em University of Wollongong}\\{\it Northfields Avenue,  Australia 2500} \\ {\it \{zf984,kwanwu\}@uow.edu.au}}
\date{}
\maketitle

\begin{abstract}
Advances in Micro-Electro-Mechanical Systems (MEMS) have revolutionized the digital age to a point where animate and inanimate objects can be used as a communication channel.  In addition, the ubiquity of mobile phones with increasing capabilities and ample resources means people are now effectively mobile sensors that can be used to sense the environment as well as data carriers.  These objects, along with their devices, form a new kind of networks that are characterized by frequent disconnections, resource constraints and unpredictable or stochastic mobility patterns.  A key underpinning in these networks is routing or data dissemination protocols that are designed specifically to handle the aforementioned characteristics.  Therefore, there is a need to review state-of-the-art routing protocols, categorize them, and compare and contrast their approaches in terms of delivery rate, resource consumption and end-to-end delay.  To this end, this paper reviews 63 unicast, multicast and coding-based routing protocols that are designed specifically to run in delay tolerant or challenged networks.  We provide an extensive qualitative comparison of all protocols, highlight their experimental setup and outline their deficiencies in terms of design and research methodology.  Apart from that, we review research that aims to exploit studies on social networks and epidemiology in order to improve routing protocol performance.  Lastly, we provide a list of future research directions.
\end{abstract}


%
%
\section{Introduction}
\begin{figure}[htb]
 \centering
  
  \rotatebox{0}{
  \includegraphics[width=4.5in]{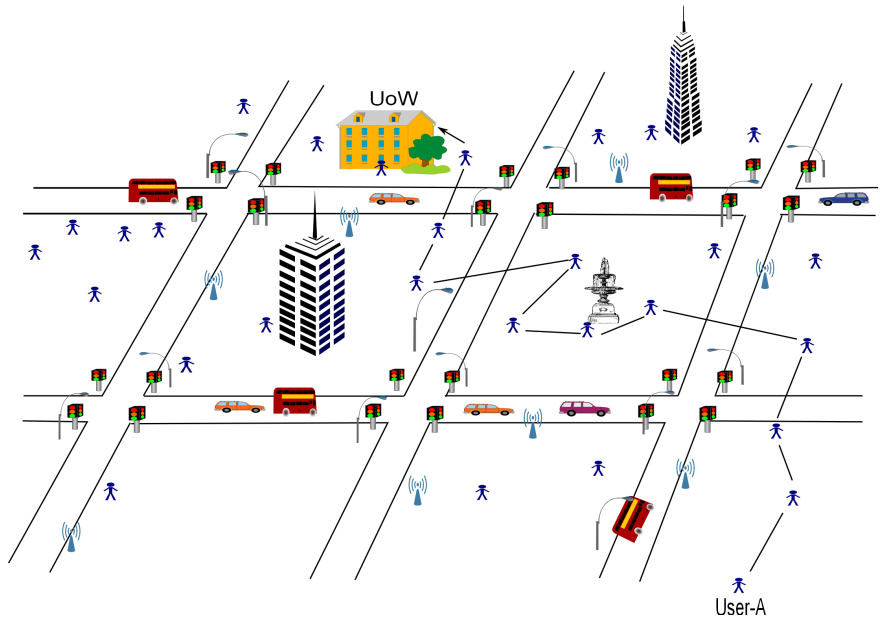}
  }
  
  \caption{An example comprising of DTNs formed by vehicles and people.} 
  \label{FIG1111}
\end{figure}

Delay or Disruption Tolerant Networks (DTNs) are characterized by long delays and intermittent connectivity.  Moreover, they may have power constraints, low and asymmetric bandwidth, and high bit-error rates \cite{FallAugust2003}\cite{RFC4838}\cite{REF58}\cite{1204759}. 
To illustrate some of these characteristics, consider forming a network using the vehicles of Figure \ref{FIG1111}.  All vehicles, e.g., buses and cars, are equipped with a radio transceiver, which allows them to communicate with each other, and also to access points, which have connectivity to the Internet and are planted strategically in different parts of the city.  In this network, all vehicles will help each other forward messages to each other and also to access points.  Given the limited transmission range, the intermittent connectivity of vehicles and location of access points, messages will experience significant delays.  Another key characteristic is the so called {\it store-carry-forward} model used to propagate messages.  That is, a vehicle may have to store and carry a message for some distance before encountering and passing the message onto another vehicle or access point.  In this regards, a key mobility pattern that can be exploited by routing protocols is the predictable mobility pattern and schedule of buses.  In addition, any routing protocols will have to consider the link capacity and duration of each connection, which is governed by channel condition and vehicle speed.

We can also form a DTN using people. This can be easily realized given the ubiquity of smart phones equipped with a plethora of sensors and transceivers.  Hence, they can be used to monitor traffic, crowd, air pollution and spread of diseases to name a few.  Unlike vehicle based DTNs, smart devices have resource constraints; e.g., limited battery life.  Moreover, people will have varying contact duration and frequency.  That is, their movement pattern will be less predictable than vehicles.
Consider User-A in Figure \ref{FIG1111} who wants to send a file to one or more students attending the University of Wollongong (UoW).  Also shown is a possible transmission path, which depends on encounters with other users of the DTN.
Inevitably, the resulting topology or path taken will be random in nature and changes with space and time.  More specifically, it is difficult to predict as it depends on the mobility patterns of people.  
Interestingly, the authors of \cite{4161914} showed that nodes/students who are attending the University of Cambridge are not always connected, and hence they experience large delays between meetings. For example, nodes/students may connect during a class, and  disconnect when the class finishes. The next class may be eight hours away. Secondly, the movement of nodes and encounter duration are random. For instance, two good friends may remain in contact for a lot longer as compared to other students. Thirdly, nodes exhibit a mobility pattern that coincides with meeting times, e.g., lectures, and path to a given classroom. 

As we can see from Table \ref{TBL1}, researchers have proposed various DTNs and applications. For example, in ZebraNet \cite{ISI:0001798}, zebras have custom sensors that track their movement patterns and locations. A device carried by a person or a mobile base station is then used to collect the said tracking data. As mobile base stations have limited communication range, zebras exchange information with other zebras until they encounter a mobile base station. Given that zebras and a mobile station rendezvous randomly, i.e., they may not meet each other for days or weeks, tracking data incurs significant delays before scientists are able to collect them. Moreover, as the movements of zebras are unpredictable, links are established intermittently and hence there are no end-to-end paths from zebras to mobile base stations. In addition, ZebraNet also has storage, bandwidth and energy constraints. Specifically, the sensors on each zebra have a lifetime of only one month, are equipped with a 1MB flash Random Access Memory (RAM) and have a data transmission rate ranging from 2.4 to 19.2 kilobits per seconds. Another wildlife example is SWIM \cite{778443}, where a sensor network is used to monitor whales. SWIM combines two kinds of nodes: (i) sensors, and (ii) infostations. The sensors are attached to whales, and the infostations are used to collect data from passing whales. In \cite{4318006}, sensors networks are used to monitor water pollution and noise level in urban areas.  

There are also a number of applications that involve special nodes. For example, DakNet \cite{1319279} uses vehicles or data ferries to provide low-cost data delivery between rural villages. In each village, a kiosk is used by villagers to store messages and send data to visiting data ferries, which then uploads the data they have collected onto the Internet. As data is carried by data ferries, it experiences a much higher delay than conventional networks. To clarify, the delays incurred by messages are affected by several data ferry parameters: (i) routes taken to reach a village, (ii) schedules, (iii) speed, (iv) number of ferries, and (v) distance between kiosks. 
In a different work, the Pollen network \cite{Glance2001429} uses humans as data ferries, where mobile devices carried by humans exchange information with each other, and/or with a fixed network.  Other projects that use data ferries include KioskNet \cite{1161127}, MotoPost \cite{1409998}, Wizzy Net \cite{REF9}, Widernet \cite{REF10}, Digital Study Hall \cite{REF11}, DigitalGreen \cite{REF11}, Body Sensors \cite{Quwaider2010}, and Digital Polyclinic \cite{REF11}. 
%

%

\begin{sidewaystable}[htbp]
\caption{\label{TBL1}A sample of DTN applications. }
{\small
\begin{center}
\begin{tabular}{|l|c|c|c|c|} \hline
\textbf{DTN Applications} & \multicolumn{1}{c|} {\textbf{Purpose}} & \multicolumn{1}{c|} {\textbf{DTN Nodes}} & \multicolumn{1}{c|} {\textbf{Delay}} & \multicolumn{1}{c|} {\textbf{Data Ferries Routes}} \\ 
\hline
SeNDT \cite{4318006} &Water pollution  &Chemical sensors, noise detectors and &Days or &Random, depending on data \\
&and noise monitor &people with mobile devices &months & requirement \\
\hline
DakNet \cite{1319279} && Coaches, motorbikes, ox carts, kiosks, &Minutes or& Semi-random, depending on the \\
&Digital communications&Access Points (APs)  &hours & transport vehicle used  \\ \cline{1-1} \cline{3-5}
KioskNet \cite{1161127} &for rural areas&Buses, people with hand-held &Hours or &Scheduled as per buses \\
& &devices, kiosks, desktop computers&day &timetable or random according\\ \cline{1-1} \cline{4-4}
MotoPost \cite{1409998} & & with a dial-up connection&Hours & to people movement \\
\hline
Pollen \cite{Glance2001429} & Personal communications & People with mobile devices & Hours or & Random, as per the \\
 & &or PDA &days & environment; e.g., in an office \\
\hline
Wizzy Net \cite{REF9}&Facilitate education & People with memory sticks &Hours or & \\
&in rural schools & & days & \\ \cline{1-1} \cline{3-3}
Digital Study Hall \cite{REF11} &  & & &Semi-random. Data ferries that visit \\ \cline{1-2} \cline{4-4}
Digital Green \cite{REF11} &Disseminating agricultural& People with DVDs and &  & villages as needed to disseminate \\
&information to rural areas& players & & information on agriculture or \\ \cline{1-2}
Digital Polyclinic \cite{REF11} &Providing healthcare &  &Days or & healthcare \\
 &information to rural areas  & &months & \\ \cline{1-3}
Widernet \cite{REF10} & To improve educational&Desktop computers with sufficient & & \\
&communication systems & storage to store web sites with rich & & \\
& in Africa &  educational contents& & \\
\hline
TrainNet \cite{Araki2010} &To transport massive amount& &Minutes or & \\
&of non real-time data over &Trains, stations &hours & Fixed, as per railway lines  \\
&large geographical areas & & & \\
\hline

\end{tabular}
\end{center}
}
\end{sidewaystable}

%
%
The aforementioned characteristics require novel routing protocols that address the following challenges:
\begin{itemize}
\item {\it Stochastic and dynamic topologies}, as nodes are mobile and can engage in various mobility patterns \cite{Daly2009}. For example, nodes may be vehicles on a freeway or wild animals roaming in a national park. These unstable topologies lead to unpredictable, uncontrolled movements, large delays and arbitrary disconnections. 

\item {\it Limited topological information}, which compounds the difficulty of finding routing metrics that accurately reflect network conditions.  A path from a source to a destination can be static or dynamic. However, as pointed out by \cite{SushantJainAugust2004}, without topology information, static routes are not suitable for dynamic topologies.  Another challenge is the lack of up to date topological information which can be used to calculate the best path to a given destination. Therefore, as will be evident later, dynamic routing protocols tend to use local metrics, such as the number of times nodes have encountered each other. 

\item {\it Variable and uncertain connection duration}. Routing protocols need to decide whether to transmit all or a subset of bundles when nodes encounter each other. For example, as zebras in ZebraNet \cite{ISI:0001798} meet for a limited time period, the routing protocol has to decide which data to forward in order to maximize delivery probability. 

\item {\it Limited resources}, which requires protocol designs to be efficient.  In other words, nodes must utilize their limited hardware resources such as CPU, memory and battery efficiently. For example, in WSNs, nodes can be located in an open environment for years before data are collected, and hence requires nodes to carefully manage their energy usage. Additionally, a good routing protocol will leverage the resources of multiple nodes. For example, nodes may choose to shift some of their stored bundles to other nodes to free up memory or to reduce transmission cost. 
\end{itemize}

To date, there are only two prior literature reviews \cite{4116780}\cite{JianShen2008}. Our paper extends these works in the following manner:
\begin{itemize}
\item We cover 42 new protocols.  Specifically, there are 18 new unicast protocols as well as 10 multicast routing protocols.  In addition, we review 14 routing approaches that incorporate either erasure or network coding in order to improve bundle delivery ratio and robustness.  Note, both prior works do not cover multicast nor coding-based routing protocols.  

\item We present extensive qualitative comparisons of all protocols. In particular, we categorize protocols according to their key feature and compare them in terms of their routing policies, resource constraints and requirements, advantages and deficiencies.   These are significant contributions as past works have only carried out a small number of comparisons using a limited set of criteria.

\item We highlight and summarize works that seek to borrow concepts from social networks and epidemiology in order to improve routing performance.    These works are particularly critical to future researchers given the ubiquity of smart phones with increasing computation power and storage.

\item Lastly, we list future works that have not been identified elsewhere.  One of which is a key limitation in regards to the research methodology used by existing works that hinders the objective and quantitative evaluation of  routing protocols.  Others include the development of multicast protocols and information retrieval systems for DTNs.
\end{itemize}

The rest of the paper is structured as follows. Section \ref{BG} first presents an overview and taxonomy of routing protocols before delving into the details of epidemic routing protocols, and their variants, in Section \ref{EPIDSEC}.  This is then followed by data ferries based protocols in Section \ref{DFSEC}.  That is, routing protocols that assume the existence of special nodes with ample resources and deterministic trajectories.  After that in Section \ref{STASEC}, we review protocols that dynamically maintain historical information of past encounters to aid its future forwarding decisions.  In Section \ref{CODER}, we review novel approaches that employ erasure or network coding.  Then in Section \ref{MSEC}, we review multicast routing protocols before reviewing research areas in social networks and epidemiology in Section \ref{ESEC}.  Finally, in Section \ref{CONC}, we present our conclusions.

\section{\label{BG}Unicast Routing Protocols}
In the forthcoming sections, we will use the term {\em bundle} as the unit of exchange between nodes. Note that the IETF/IRTF defines bundle as the metadata that is used for wrapping information from other layers and to compress them into a data block.  Moreover, DTN routing protocols are independent of the delay tolerant networking architecture presented in \cite{RFC4838}.

\subsection{Overview}
Current DTN routing protocols can be divided into three categories: (i) epidemic, (ii) data ferries and (iii) statistical. Table \ref{TAB2} summarizes the key characteristics of each category. We see that the three categories have different features, advantages and disadvantages. 

\begin{figure}[htb]
 \centering
  
  \rotatebox{0}{
  \includegraphics[width=4.5in]{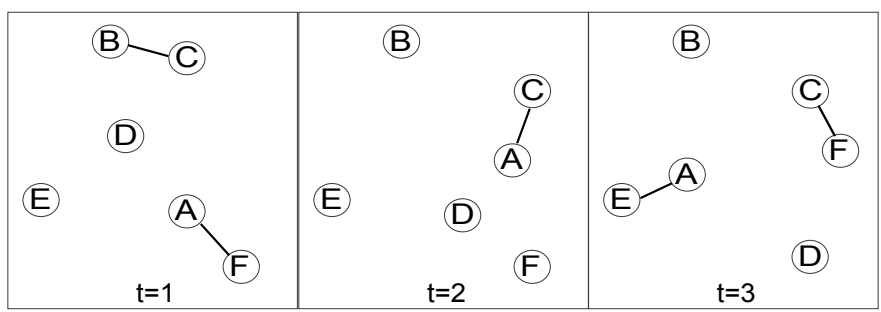}
  }
  
  \caption{A DTN with homogeneous nodes.  A solid line shows connectivity between nodes.} 
  \label{FIGSTOCEG1}
\end{figure}

Nodes in category (i) are assumed to have uniform resource and movement patterns.  Moreover, they cooperate to route bundles for their neighbors.  For example, in Figure \ref{FIGSTOCEG1}, at $t=1$, assume node B wants to forward a bundle to node-E.  One approach that node-B can adopt is to simply forward the bundle to any nodes it encounters; i.e., floods the bundle as widely as possible.  We see that the bundle arriving at node-E following the path B--C--A--E at $t=3$.  Unlike conventional networks, this path, however, is likely to change in subsequent bundle transmissions.  In this respect, the key research objective, as we discussed in Section \ref{EPIDSEC}, is to design an efficient flooding based protocol that meets the following goals: a) message delivery rate is maximized, b) message latency is minimized, and c) the total resource, especially memory or energy expenditure, consumed is minimal. In general, epidemic routing protocols have low delays, but high resource consumption.  Hence, designing a buffer management policy that balances resource consumption and delivery ratio is a fundamental problem.  Moreover, said policy needs to purge staled bundles reliably.  Otherwise, prematurely removing bundles may have a negative impact on delivery ratio.

\begin{figure}[htb]
 \centering
  
  \rotatebox{0}{
  \includegraphics[width=4.5in]{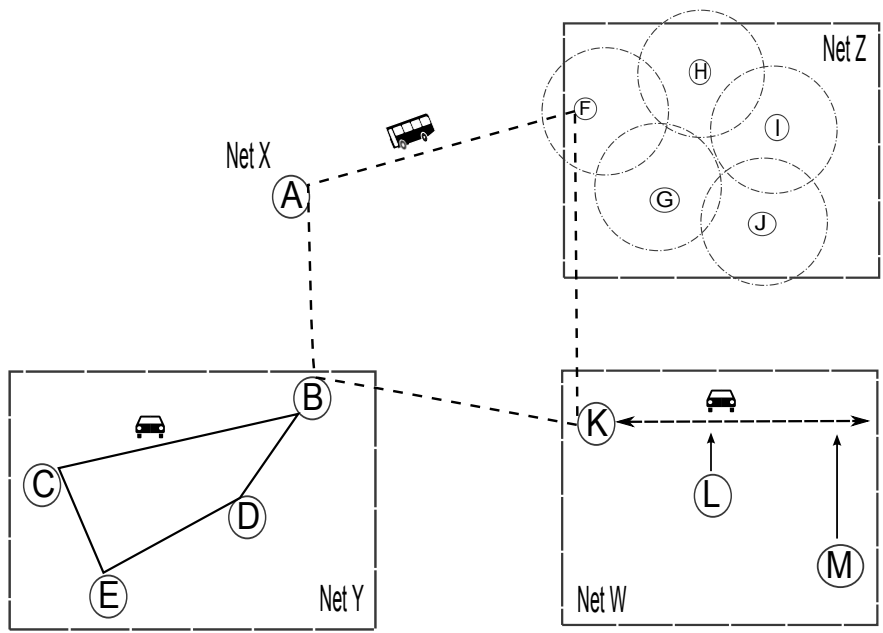}
  }
  
  \caption{Example of DF based DTNs.} 
  \label{FIGDFEG1}
\end{figure}

Routing protocols in category (ii) take advantage of resource rich, mobile nodes called Data Ferries (DFs), which act as a communication channel between nodes or disparate networks.  These networks could be located on different planets or represent rural villages.   
Examples of DFs based DTNs are shown in Figure \ref{FIGDFEG1}.  We see four networks that are serviced by a DF (bus), which tours each network periodically.   Note that one can reduce delays further by adding more DFs or buses.  Each network, e.g., Net-Y, can be serviced by an independent DF.  That is, instead of networks, a DF (car) is tasked with collecting data from nodes directly.  Apart from that, nodes can rely on wireless communications, i.e., Net-Z, and only use a DF for inter-cluster communications; see Section \ref{DF5} for more details.
The figure also illustrates two kinds of data ferry movements: active and passive. In the former, data ferries actively approach source or destination nodes; e.g., the bus.  In the latter model, nodes intentionally move toward data ferries, as illustrated by nodes in ``Net-W''.   In both types of data ferries, they have a predictable schedule which a routing protocol can then exploit to provide some form of Quality of Service (QoS) guarantees.  
%
%
%


The last category, i.e., (iii), of routing protocols utilizes statistical methods to avoid arbitrary flooding. The key assumption is that nodes in a DTN will always encounter one another.  Moreover, nodes are homogeneous in that they have similar resources. Hence, each node can compile a set of statistics or metrics regarding its rendezvous time with other nodes. For example, in Figure \ref{FIGSTOCEG1}, we see that node A has a connectivity.  That is, it meets other nodes frequently, and thereby, making it ideal as a bundle carrier.  As we will elaborate in Section \ref{STASEC}, nodes may  forward a bundle to a neighbor based on statistical information such as next hop forwarding probability. Other metrics include the number of times a node has encountered a given node, and the duration in which a node remains connected to a given neighbor.  Moreover, routing protocols may consider the storage capacity, energy, bandwidth and/or type of nodes.  As a result, routing protocols in this category result in nodes with lower buffer occupancy consumption.  The downside, however, is gathering invariant properties of a DTN which a node can then exploit to forward its bundles.

In the next section, we first outline epidemic routing protocols and their optimizations related to buffer management.  Then in Section \ref{DFSEC}, we survey works that exploit and control special nodes or DFs.  Lastly, in Section \ref{STASEC}, we expound works that exploit mobility patterns.
%
%
\begin{sidewaystable}[htbp]
{\scriptsize \caption{\label{TAB2}A comparison between DTNs routing protocols categories}}
{\small
\begin{center}
\begin{tabular}{|l|l|l|l|} \hline
& \multicolumn{1}{c|} {\textbf{Epidemic}} & \multicolumn{1}{c|} {\textbf{Data Ferries}} & \multicolumn{1}{c|} {\textbf{Statistical}} \\
\hline
{\textbf{Forwarding Strategy}} & Pure or limited Flooding  & Reactive/Proactive & Proactive \\   
\hline
{\textbf{Node Types}}             & Homogeneous & Heterogeneous & Homogeneous \\
\hline
{\textbf{Mobility}}               & Random      & Controlled  & Semi-random \\
\hline
{\textbf{Delay}}                  & Lowest      & Highest       & Moderate \\
\hline
{\textbf{Bundle Duplication}}  & Every node encounter & Upon encountering a DF & Only to neighbors that meet a given \\
                               &                      &                        & history criterion \\
\hline
{\textbf{Energy Expenditure}}     & Highest     & Lowest        & Moderate \\
\hline
{\textbf{Maintain Encounter}}  & No        & Partially           & Yes \\
{\textbf{Information}}         & & & \\
\hline
{\textbf{Location Information}}   & No          & Yes           & No \\
\hline
{\textbf{Complexity}}             & Simple      & Moderate           & Highest \\
\hline
{\textbf{Remarks}}             & (i) Strong theoretical support, (ii)  & (i) Data ferries have unlimited storage, & (i) Changes in topological 
                                                                                                         properties  \\
                               &  flooding of bundles increases buffer  & (ii) finding a tour that minimizes delay & affect convergence time, (ii) nodes    \\
                               
                               &  occupancy level and energy  expenditure, & is NP hard, (iii) buffer occupancy  &  are required to calculate or record     \\
                                  & (iii)  buffer management policies trade & is dependent on DFs' tour & statistics or historical data at every \\                                                                 
                                 &  off delivery ratio and buffer occupancy & length  &  encounter \\                                 
                                 &  level.                                  &   &  \\                                
\hline
\end{tabular}
\end{center}
}
\end{sidewaystable}

\subsection{\label{EPIDSEC}Epidemic}
\subsubsection{Pure}
Vahdat et al. \cite{Vahdat2002} proposed the first epidemic routing protocol for DTNs, which is based on an epidemic algorithm originally developed for updating databases \cite{43922}. Each node maintains a bundle summary vector describing each bundle's destination, length and hop count. Whenever two nodes encounter each other, they begin an anti-entropy session, where they compare their bundle summary vector to ascertain missing bundles. Nodes stop their anti-entropy session when they have the same bundle summary vector; i.e., same set of bundles. Each bundle also contains a hop count that corresponds to its priorities, and is also used to constrain flooding. Apart from that, nodes allocate a dedicated buffer space for non-local bundles. 
In their experiments, the authors compared the delivery rate when nodes are allocated varying buffer sizes, and when bundles have different hop counts. Their results show that using an infinite buffer size results in the fastest message delivery, and achieves 100\% bundle delivery with an average time of 147.3 seconds. With a buffer size of 1MB, 500KB and 200KB, they only observe up to 0.4\% and 10\% degradation in average delivery rate and maximum delivery times respectively. Interestingly, when the buffer size shrinks to 2.5\% of its original capacity, the delivery rate reduces to 79.7\%, and only 29.3\% when nodes have a 10KB buffer size. Their results also show that protocol delivery rate is affected by bundles hop count. When hop count is changed from eight to one, delivery rate drops from 100\% to 80\%.  In general, their results show that buffer size has a non-negligible impact on bundle delivery ratio.  
Although these results confirm epidemic routing has good performance in DTNs, it suffers from the following problems. Firstly, in scenarios where nodes have a large summary vector or number of bundles, a short contact period may prevent nodes from exchanging their summary vector successfully and complete an anti-entropy session. Secondly, nodes do not preferentially discard bundles during congestion.  Thirdly, its performance is limited when nodes have memory constraints. 

\subsubsection{Optimizations}
To date, researchers have proposed two main ways to reduce the memory consumption of nodes: anti-packets, limited copies, and metrics.  

\subsubsection*{Anti-packets}
An anti-packet is generated by a destination node once it has received a bundle successfully.  In addition, anti-packets are used to eliminate duplicated bundles.  That is, each anti-packet is paired with a bundle, analogous to ``infection and vaccination'' pair in epidemiology.  Note, anti-packet is analogous to vaccination \cite{4530735}, \cite{4428685}, immunity \cite{4557809} or cure \cite{4483165}.  Hence, upon receiving an anti-packet, nodes check their bundle list and delete the corresponding bundle.  

The $(p,q)$-epidemic with vaccination routing protocol or PQERPV \cite{4530735} transmits bundles with varying probability and avoids duplicating bundles. Specifically, nodes store bundles from their encounters with probability $p$ and $q$, where $p=q=0$ means nodes do not store any bundles, and $p=q=1$ indicates nodes are to receive all bundles. Nodes that are neither sources nor destinations are called relay nodes. Moreover, nodes that carry bundles on behalf of other nodes are termed ``infected relay nodes''. The value of $p$ and $q$ is set according to node types.  Specifically, relay nodes store bundles from sources with probability $q$ and store bundles from infected relay nodes with probability $p$.  Destination nodes initiate a vaccination process once they have received a bundle correctly.  Specifically, they flood anti-packets to delete the corresponding bundle buffered at relay nodes. This protocol has the following limitations. Firstly, the flooding of anti-packets does not consider resource consumption. Secondly, the speed in which source nodes send new bundles is governed by the dissemination of anti-packets.  In other words, source nodes will not send new bundles until they have the corresponding anti-packets for bundles sent previously.  In effort to reduce resource consumption, in \cite{KhaledA.Harras2005}, the authors combine PQERPV with an additional metric called Time-to-Live (TTL) to control and eliminate delivered bundles.  Their results show that when TTL is set to 10 minutes, nodes store on average 27\% of transmitted bundles as compared to 19\% when their TTL value is set to five minutes.  This is because nodes are more likely to remove expired bundles.

Another protocol that utilizes anti-packets is Epidemic Routing Protocol with Immunity \cite{4557809}\cite{Mundur2}. This protocol delivers and drops bundles according to two lists: m-list and i-list. m-list is similar to the summary bundle vector in epidemic routing protocol \cite{Vahdat2002}, where it records a bundle's ID and destination. i-list maintains the bundle IDs that have arrived at their respective destination. When two nodes encounter each other, they combine their i-list and exchange those bundles that are not in their i-list. Each bundle that successfully arrives at its destination triggers nodes to update their i-list.  A key concern, however, is that the size of i-list will increase following the successful delivery of bundles.  Moreover, the authors did not specify any i-list management policies. 

Scheme for Epidemic Routing with Active Curing (SERAC) \cite{4483165} proposes a faster anti-packet transmission and efficient buffer management scheme. The main idea, called active curing, is to prioritize the transmission of acknowledgment (ACK) messages, and thereby propagate them quicker throughout the network. Additionally, SERAC recalculates a new route when forwarding ACKs so that they follow the ``best'' path given the current network state.  This, however, consumes more resources in terms of memory and CPU computation.   Besides that, to minimize the size of ACKs, SERAC uses two bytes to represent the sequence of bundles that has arrived at a destination. However, when bundles are fragmented or small in size, these two bytes overhead will be significant. 

\subsubsection*{Limited Copies}
Spyropoulos et al. \cite{MCOPY1}\cite{MCOPY2} propose the Spray and Wait routing protocol.  Its routing process can be split into two parts: (i) spray phase, where $L$ copies of a bundle are initially forwarded by a source to $L$ neighbors, and (ii) in the wait phase, these $L$ neighbors relay a copy of the message only when they encounter the destination.  The same authors also introduce an improvement called Binary Spray and Wait scheme, in which each node transmits half of their bundles they have to any encountered nodes.  For example, a source node with $L=10$ will transmit five bundles to another node-A, and keeps five bundles for itself.   This process is then repeated for any nodes that the source and node-A meet in the future.  Their experiments involving 100 nodes show that with $L$ increasing from five to 20 the delivery delay decreases by approximately 42\%.  Under the same condition, Binary Spray and Wait has a higher performance, where delivery delay ranges from 3500 to 1500 seconds.   The main limitation with this protocol is that a maximum of two hops is used to deliver bundles.  Hence, in large DTNs, a bundle may incur a significant delay as it can only be delivered when a relay or source node encounters the destination. 
%
In \cite{Spray2010}, Bulut et al. consider how bundle copies are distributed to relays given a time constraint.  The main approach is to have a number of periods, each with increasing ``urgency'' corresponding to a bundle's deadline.  Initially, a source sends out a small number of copies, and waits for an acknowledgment.  If delivery fails, the source {\it sprays} additional copies to nodes that have yet to receive a copy of the bundle.   Hence, with each passing period, more copies are generated to ensure a bundle is delivered. The authors show via analytical and simulation studies that multiple periods reduce the number of bundle copies required to meet a given deadline.  Their work, however, assumes acknowledgments are forthcoming in each period to facilitate bundle transmission.  In addition, acknowledgments are forwarded using epidemic routing, which incurs high overheads.

%
%
Energy is an important issue for DTNs comprising of battery constrained devices; e.g., sensors attached to animals \cite{ISI:0001798}.  To this end, Li et al. \cite{EffOpp2010} and Altman et al. \cite{EffAlt2009} study energy efficient forwarding policies for these types of DTNs.  That is, limit bundle copies so that nodes incur minimal energy expenditure associated with transmission and reception.  In \cite{EffOpp2010}, the authors consider a two-hop forwarding model, whereas the latter work also considers probabilistic epidemic forwarding.  In both works, the goal is to seek to design policies that improve bundle delivery whilst adhering to a given energy budget or bounded transmission times.  The key control parameter is the probability of transmission.  For example, in two-hop forwarding, a source forwards a message to another node at time $t$ with probability $p(t)$. Using an extensive analytical framework, Li et al. showed that setting $p(t)$ according to the following policy to be optimal in terms of energy efficiency: given a time threshold $t_0$, set $p(t) = 1$ if $t \le t_0$, and $p(t)=0$ otherwise.

\subsubsection*{Metrics}
In this optimization, nodes compute a metric that reflects the current network state, such as the number of times a node has encountered another node or contact duration, in order to evaluate a neighbor's ability in delivering a bundle successfully.  

James et al. \cite{962117} developed an epidemic routing protocol that uses a buffer management strategy called Drop-Least-Encountered (DLE). The authors utilize Encountered Count (EC) as a metric to decide which bundles to drop when a node's buffer is full. Specifically, nodes drop bundles with the smallest EC. Apart from that, the value of EC degrades with time. For example, if two nodes meet only once, their EC value will decline from one to zero gradually. Additionally, nodes can also learn the EC metric of a neighboring node's past encounters. Consider node A, B and C. When node A encounters node C after meeting node B, node C can also learn the EC metric of node B and A.  However, this protocol still uses flooding to transmit bundles, and it does not consider bundles priority. Consequently, high priority bundles may be dropped when a node's buffer is full. 

Prioritized Epidemic (PREP) routing protocol \cite{1247707} gives preference to bundles according to a cost value that is computed according to their respective destination, source and expiry time. In addition, PREP maintains a high replication density when bundles approach their destination. It consists of two main components: topology awareness and bundle drop/transmit processing. The topology awareness component updates a link metric called Average Availability (AA), which is defined as $\frac{T_{up}}{T_i}$, where $T_{up}$ is the total time when the link is up, $T_i$ is the time when the link is available. A bigger AA value means a link has a higher utility and stability. PREP marks bundles that are further away from their destination as low priority. This means nodes maintain a high bundle density as bundles approach their destinations. The main problem with PREP is that in highly dynamic topologies, nodes will announce updated AA frequently, which slows route convergence. 

\subsubsection{Discussion}
Table \ref{TAB3} summarizes epidemic based routing protocols.  We see that since Vahdat et al. \cite{Vahdat2002}'s seminal work, a number of works have been proposed to address many of its limitations.  They range from the use of probabilistic metrics to ``intelligently'' forward bundles to the removal of packets via anti-packets or expiration time.  The main limitations of these protocols are that they assume all nodes  have the same capabilities or properties, and consume a significant amount of resources.  In the next section, we outline protocols that exploit resource rich nodes that have specific trajectories to help propagate bundles.

%
%
%
\begin{sidewaystable}[htbp]
{\scriptsize \caption{\label{TAB3}A summary of epidemic routing protocols.}}
{\small
\begin{center}
\begin{tabular}{|l|l|l|} 
\hline
{\textbf{Protocol}} &{\textbf{Techniques to Minimize Flooding}} & {\textbf{Key Feature}}\\
\hline
Pure Epidemic \cite{Vahdat2002} & Transmit randomly & Duplicate bundles until they are delivered\\
\hline
&Transmit according to probability & Use of anti-packets to erase duplicated bundles \\
PQERPV \cite{4530735} & $p$ and $q$ & \\
\hline
& & Use of anti-packets to erase duplicated bundles \\
Epidemic with & & \\
Immunity \cite{4557809} & & \\ \cline{1-1} \cline{3-3}
& Transmit bundles randomly & Prioritized ACK delivery \\
SERAC \cite{4483165} & & \\ \cline{1-1} \cline{3-3}
& & Combines buffer management with encounter times \\
Epidemic with DLE \cite{962117} & &  \\
\hline
PREP \cite{1247707} & Relies on the Open Shortest Path & Paths have a cost value, and bundles are forwarded along the path with the\\
& Forwarding \cite{REF66} routing protocol & the least cost \\
& to compute a least cost path &  \\
\hline
Epidemic with TTL \cite{KhaledA.Harras2005} & Transmit according to a probability value & Bundles have an expiration time that is determined by their TTL value \\
\hline
Spray and Wait \cite{MCOPY1}\cite{MCOPY2} & Limited copies. & Bounded bundle copies.  Two-hops delivery as relays only forward bundles \\
                                          &                 & directly to a destination.\\
\hline
Stochastic \cite{EffOpp2010}\cite{EffAlt2009} & Dynamic adjustment of forwarding & Considers the energy expenditure of bundle transmission and replication.  \\
                                              &  probabilities & \\ 
\hline
\end{tabular}
\end{center}
}
\end{sidewaystable}

\subsection{\label{DFSEC}Data Ferries}
The second category of routing protocols involves the use of Data Ferries (DFs).  To date, there are 15 protocols that employ DFs.  In general, any nodes can be DFs; viz. hand-held devices, vehicles, animals, and satellites.  DFs follow a predictable or unpredictable pattern. For example, buses move periodically but with a period that ranges from a few minutes to several hours.  On the other hand, wild animals have stochastic movement patterns.  Lastly, a DTN can have one to hundreds of DFs \cite{1204352} \cite{1498365}.
  
We will first review works that investigate the general benefits of DFs in Section \ref{DF1}.  Following that, in Section \ref{DF2}, we present works that make use of vehicles as DFs.  Then in Section \ref{DF3}, we survey works that use DFs to reduce the energy consumption of nodes.  After that, in Section \ref{DF4}, we outline the use of DFs in deep space.  Lastly, Section \ref{DF5} reviews works that consider group of nodes, each of which is served by one or more DFs.

\subsubsection{\label{DF1}General}
The Message Ferrying (MF) \cite{989483} scheme classifies MF movement and routing into two categories: (i) Node-Initiated MF (NIMF), in which DFs move around a deployed area according to known routes and communicate with other nodes they meet, and (ii) Ferry-Initiated MF (FIMF), in which DFs move pro-actively to meet nodes. The authors evaluate the performance of MF according to their data delivery rate and energy consumption. Their results demonstrate that MF has a higher delivery rate as compared to epidemic routing protocol, and advantageously consume less energy.  
Moreover, MF transmits 30 times more data per Joule of energy than epidemic routing protocol.  

In \cite{1204352}, the authors investigate the relationship between the number of DFs and routing performance.  Specifically, they report on the use of a single and multiple DFs. The used the following equation to calculate the delay incurred by a DF to collect data \cite{989483}: 
\begin{equation}
D=\frac{\sum_{0<i,j\le n} w_{ij}d_{ij}}{\sum_{0<i,j\le n} w_{ij}} 
\label{EQU1}
\end{equation}
where $n$ denotes the number of stationary nodes, $w$ is the data transmission rate, and $d$ is the delivery delay of each data fragment. Note that the problem of computing the DF path that minimizes delay is effectively the well known traveling salesman problem. 
Their approach, however, has only been applied in DTNs with static nodes. 

In their subsequent work \cite{1498365}, the authors investigate the use of multiple DFs, and the following routing schemes: single route (SIRA), multiple routes (MURA), relaying between ferries indirectly (NRA), and directly (FRA).  The DFs in SIRA travel on the same trajectory, but are on different trajectories in MURA. The difference between NRA and FRA is whether stationary nodes are used as relay nodes between DFs.  Their experimental results show that MURA is able to take advantage of routes with the least delay.  
Also, increasing the number of DFs is also beneficial for reducing the buffer consumption of DFs. Their experiment only considers the scenario where data ferries operate in semi-static topology.  However, in high mobility scenarios, data ferries are required to track nodes, and coordinate the delivery process with each other -- a task that incurs high signaling overheads.  

\subsubsection{\label{DF2}Vehicular Networks}
The authors of \cite{1023884} propose Mobility-Centric Data Dissemination Algorithm for Vehicle Network (MDDV), which they then apply to Vehicle-to-Vehicle (V2V) networks with the following characteristics: (1) predictable and high mobility, (2) dynamic and rapidly changing topology, (3) constrained and mostly one-direction movements, (4) potentially large-scale, (5) frequent disconnections, (6) vehicles that are not completely reliable, and (7) no significant power drain.  MDDV exploits the mobility of vehicles to deliver bundles and combines opportunistic and trajectory forwarding as vehicles only encounter each other occasionally, move along streets or roads, and are aware of their geographical location -- as provided by the Global Position System (GPS).  Nodes using MDDV send bundles to neighboring nodes that are geographically closest to the region that contains the destination.  These nodes then flood the bundles upon entering the designated region. This research has two key issues. Firstly, there is no route recovery mechanism. That is, the authors have not considered the case when bundles fail to be delivered due to data ferries running out of patrol. Secondly, the use of flooding is prohibitive when nodes have a high mobility. 

Another routing protocol is Meeting-Visit (MV) \cite{1497909}.  Nodes build a geographical information profile whenever they meet peers and visit different geographical locations.  MV works as follows. When two nodes meet each other, they will first exchange a list of bundles they are carrying as well as their corresponding destination.  In addition, they will annotate these bundles with a delivery probability.  After that, the pair of nodes will sort bundles according to their delivery probability. Bundles that are headed in the opposite direction will be allocated the lowest probability.  They then select the top $n$ bundles with the highest delivery probability. The delivery probability is calculated as follows: 
\begin{equation}
P^k_n(i)=1-\prod\limits^N_{j=1}(1-m_{jk}P^j_{n-1}(i))
\label{EQU2}
\end{equation}
where $P^k_n (i)$ denotes the success probability that node $k$ transmit $n$ bundles, $N$ is the number of nodes, $m_{jk}$ represents the probability that node $j$ and $k$ visit the same location simultaneously. 

MV also introduces four kinds of controllers: (i) total bandwidth, which chooses the peer which has not encountered other nodes for the longest period of time, (ii) unique bandwidth, which chooses the peer that has the largest number of messages not present anywhere else in the network, (iii) delivery latency, which chooses the peer whose average delivery time is the largest, (iv) peer latency, which chooses the location least recently visited by a peer. A multi-objective controller then allocates a different priority to the aforementioned controllers. Their experiments show that although nodes experience higher latencies, MV can deliver messages with a high success ratio -- i.e., 83\% of the maximum achievable delivery rate with minimal duplicated bundles. However, their experiments show that MV has marginal performance as delivery rate does not improve significantly with increasing node density. 

Zarafshan-Araki et al. \cite{Araki2010} propose TrainNet, a system that uses trains as DFs to transport massive amount of data between stations on a single railway track.   A key problem observed is storage optimization, where the storage capacity of hard disks being loaded and unloaded from each station becomes a bottleneck given that trains only stop for a relatively short period of time at each station.  They propose  four scheduling algorithms to manage the said hard disks: (i) First Come First Serve (FCFS), in which data is loaded onto trains according to their arrival time; (ii) Local Max-Min Fair Algorithm (LMMF), in which the farthest stations are allocated more capacities on train's hard disks; (iii) Global Max-Min Fair (GMMF), in which equal storage capacities are allocated to downstream stations; (iv)  Weighted Global Max-Min Fair (WGMMF), in which data is prioritized by weight.  Their results show that WGMMF has the best hard disk utilization. 
Note that WGMMF also has the highest average throughput, which is significantly more than GMMF. This work suffers from the following limitations: (i) intermediate stations cannot be used for storing data, (ii) the protocol does not support train-to-train communication, and (iii) the work only considers trains travelling in a single railway track.   

\subsubsection{\label{DF3}Energy}
Another key issue in DTNs is the energy constraint of nodes, which has a significant impact on a node's ability to deliver bundles. Zhu et al. \cite{4685156} present two algorithms, called Least Energy Tree (LET) and Minimum Hop Tree (MHT), to provide energy efficient message ferrying in wireless sensor networks.  Both LET and MHT are based on the spanning-tree algorithm.  The algorithms construct a spanning tree rooted at each node.  For LET, each branch of the tree is allocated a weight corresponding to the energy needed to deliver a bundle.  On the other hand, all branches are set to one for MHT.  Note that, the spanning tree needs to be recomputed if there are topological changes.
Their results show that LET consume slightly less energy than MHT.
%
In addition, increasing the number of nodes also has an impact on the energy consumed by mobile nodes. 

The authors of \cite{1424832} consider conserving the energy of DFs.  They propose ferry replacement protocols where different nodes take turns to be DFs.  The first protocol requires the current ferry to designate a successor that takes over its functionality upon failure.  Specifically, it appoints the first node that it meets that has a higher capability than a given threshold.  In the second protocol, nodes conduct an election.  Each node computes a backoff delay according to their capability.  Nodes with a shorter backoff delay will become the next DF. Their results show 15\% reduction in overheads when data ferries have a nominated successor.  However, in the second protocol, nodes spend more than 1000 seconds electing the next ferry, which delays bundle delivery by at least 2000 seconds. However, the protocol can fail to designate or elect a successor in two scenarios: (1) when all DF candidates have the same backoff delay, and (2) the capability of the current DF falling below a given threshold before it is able to appoint a new DF. 

In \cite{1203354}, DFs, also known as MULEs, are used to provide cost-effective connectivity in sparse sensor networks and to reduce the power requirement of sensors. The authors propose a three tier architecture comprising of access points (APs), MULEs and sensor nodes.  APs are connected to the Internet, and are used for storing and analyzing data from sensor nodes. MULEs have large storage capacities, renewable power and connect asynchronously to the APs and sensor nodes.  In addition, they can communicate with each other.  The authors proved that i) a high density of MULEs improves system robustness, and ii) sensor buffer requirements are inversely proportional to the number of MULE nodes. Specifically, when the authors increase the percentage of MULEs from 0.1\% to 10\%, each with an infinite-buffer, they show a 99.5\% reduction in buffer consumption.    Nevertheless, to guarantee 100\% delivery, their scheme requires duplicated bundles.

\subsubsection{\label{DF4}Space Communications}
The advantages of using DFs in deep space networks are as follows \cite{4374103}.  Firstly, DFs cut down the long distance between two planets into relatively shorter ones, which allows each segment of the route to have a high data rate.  Secondly, all relays share the energy expenditure of transmitting data on a given route.  Lastly, DTNs provide better connectivity as other planetary objects can be used as relays.  For example, one can set the moon to be a DF which helps relay messages when Mars is not visible from Earth.  In particular, these messages can contain commands that control a spacecraft. 

Henceforth, researchers have devised numerous routing protocols for Inter-Planetary Network (IPN). The Interrogation-Based Relay Routing (IBRR) \cite{1189163} protocol is designed for free space data transmission between Low Earth Orbit (LEO) satellites. When two satellites are in the vicinity of each other, they engage in an interrogation process to exchange orbital information and routing tables. Each satellite will then decide the next best hop according to the following metrics:
\begin{itemize}
\item Spatial location and orbital information
\item Link bandwidth
\item Relative velocity/mobility
\item Vicinity of a satellite to other satellites and ground stations
\item Memory capacity
\item Data rate
\end{itemize}

IBRR also provides multipath routing. If there are more than one disjoint route to the destination, the source stores all these routes in its cache and selects the best available route.  Upon a path failure, it selects an alternate path and routing resumes. However, there are several problems: (a) the authors have not compared the impact of the aforementioned metrics. They assume that if one node has a lower bandwidth but with a higher mobility, nodes will not be able to decide which metric is best, (b) the protocol does not provide a route when a destination is more than one hop away, and (c) each satellite does not consider whether a neighbor has any success in delivering bundles to a given destination

\subsubsection{\label{DF5}Inter-Cluster Communications}
To date, there are several cluster-based routing protocols.  In \cite{1247710}, nodes in a cluster employ the Destination Sequenced Distance Vector (DSDV) \cite{190336} routing protocol, whereas a single DF is used to provide a communication channel between clusters.  Each cluster has multiple gateway nodes that transmit and receive bundles to/from a DF.  Nodes in a cluster send their bundles to gateway nodes using one of the following transmission policies: (a) {\it random} -- nodes uniformly pick and transmit their bundles to a gateway, (b) {\it proportional} -- nodes send bundles to the gateway with the highest storage capacity, and (c) {\it nearest} -- nodes choose the gateway with the shortest hop count.  Upon meeting a gateway, a DF first transmits any outgoing traffic before accepting incoming traffic or vice-versa.  Alternatively, it could transmit bundles in a round-robin manner.

Mobile Relay Protocol (MRP) \cite{1035720} combines the advantages of traditional and DTN routing protocols.  MRP assumes that every node is mobile, and nodes are connected by a traditional routing protocol, i.e., DSDV \cite{190336}, and also form a virtual cluster. Relay nodes store bundles until they can be transmitted; i.e., they meet a node with a valid route to the destination. If traditional routing fails to find a route, bundles are handed over to the DTN layer.  Once relay nodes find a valid route, it reverts back to conventional IP routing.  The authors evaluated the delivery rate and latency of MRP using three mobility models: (i) {\it random} -- each node selects its destination randomly within an allowed area, (ii) {\it soccer player} -- a node has a higher probability of picking a nearby destination, and (iii) {\it homing pigeon} -- each node has a ``home'' location, and its speed is distributed uniformly between 0 and 1 m/s, and nodes choose a random point as the destination with a probability of $\frac{1}{r^2}$, where $r$ is the distance from the ``home'' to a given destination point, before moving  back to its ``home'' base.  In each mobility model, MRP is able to deliver over 95\% of the bundles.  Nodes experience the best latency when they move according to the homing pigeon model. In addition, they show that MRP is unaffected by node density.  There are two problems with MRP.  Firstly, transmission between clusters is inefficient because bundles are only handed to their respective destination when a DF arrives at the corresponding cluster.  Hence, bundles experience variable delays.  Secondly, as MRP relies on DSDV, it is not suitable for sparse DTNs that have no contemporaneous paths between nodes. 

In \cite{1139396}, the authors propose a protocol that consists of the following messengers/DFs: regional and independent.  A regional messenger only carries bundles for the region it belongs to, while the latter one is used for delivering bundles to all regions. The authors also propose the following strategies: (i) {\it periodic} -- DFs depart regions according to pre-determined schedules, (ii) {\it on-demand} -- DFs will leave a region as soon as any nodes require transmission, and (iii) storage-based -- DFs leave its region once its buffer is full. Their results show that in terms of average delay, on-demand outperforms storage-based by 50\%. On the other hand, the storage-based scheme performs best in terms of transmission cost.  That is, in the storage-based scheme, DFs travel $\frac{1}{10}$$^{th}$ of the trips required by the on-demand scheme. However, the problem is that the delivery strategies of DFs are fixed and are not adaptive to changing network parameters.  We postulate that combining all three strategies to be a worthwhile approach. That is, a DF can modify its movement based on memory utilization, transmission requirements from nodes, and allocated paths.       

The Context-Aware Routing (CAR) \cite{1443499} protocol elects nodes as DFs if they have a high number of connections and energy within a cloud or a given geographical area.  These DFs are then responsible for ferrying messages between clouds. Their experiments compare CAR's performance with pure flooding and epidemic \cite{Vahdat2002}. Their results show that CAR achieves a 70\% delivery rate -- assuming nodes move according to the random way-point model.  This is 10\% higher than pure flooding, but 10\% lower than epidemic. However, their experiments only involve one cloud.  Therefore, the performance of CAR in multi-clouds scenarios is unclear. Moreover, the authors did not specify any protocols for communications between DFs. 

\subsubsection{Discussion}
Table \ref{TAB4} summarizes the key features of the aforementioned DF-based routing protocols. These protocols can be characterized by (i) the type of nodes that are designated as DFs, (ii) routes taken by DFs, and (iii) the number of DFs.  In particular, we can see various DF types; e.g., buses and satellites.  Apart from that, DFs have a variety of mobility patterns, and have ample resources.  Moreover, depending on the application, multiple DFs can be deployed to promote better energy efficiency, delivery rate, and reliability.   The fundamental assumption of these protocols is that there exists one or more nodes/DFs with well known properties.  However, this assumption is not valid in DTNs where all nodes have pseudo-random movements; e.g., humans.  In the next section, we present protocols that dynamically learn the invariant properties of a given network in order to route bundles. 
%

%
%
\begin{sidewaystable}[htbp]
{\scriptsize \caption{\label{TAB4}A summary of data ferrying routing protocols.}} 
{\small
\begin{center}
\begin{tabular}{|l|l|l|c|c|l|} 
\hline
\multicolumn{1}{|c|}{\textbf{Protocol}} & \multicolumn{1}{c|} {\textbf{DF Type}} & \multicolumn{1}{c|} {\textbf{Nodes Movement}} & \multicolumn{1}{c|} {\textbf{Ferry type (NI/FI)}} & \multicolumn{1}{c|} {\textbf{Number of DFs}} & \multicolumn{1}{c|} {\textbf{Target Applications}} \\
\hline
MF \cite{989483} & &Random waypoint &&& \\ \cline{1-1} \cline{3-3}
MF with single &&& & 1 & \\
ferrying node \cite{1204352} & &Stationary nodes &&& \\ \cline{1-1} \cline{5-5}
MF with multiple & &with deterministic &FI && \\
ferrying nodes \cite{1498365} & & DFs movement && & Generic \\
& Designated nodes &&& &  \\ \cline{1-1} \cline{3-4}
 & & Nodes periodically & & & \\
MV \cite{1497909} & & visit known & FI && \\
& & locations & &$>1$ & Exchange of personal data \\ \cline{1-4}
&Nodes with & On demand & &  & \\
& large storage & movements between &FI & & \\
MULE \cite{1203354} & capacities and & sensors and APs & & & \\
& renewable energy &between sensors & & & \\
& & & & & Tracking, and \\ \cline{1-5}
DF and body & & & & &monitoring \\
sensor \cite{Quwaider2010} & & As per human & & & \\
& &movement & & & \\
&Sensor & &FI & & \\ \cline{1-1} \cline{3-3}
Energy-efficient & &As dictated by a & & $\ge 1$ & \\
MF \cite{4685156} & &minimum & & & \\
& &spanning tree & & & \\ \cline{1-4} \cline{6-6}
MDDV \cite{1023884} &Vehicles &According to  &Both & &V2V \\
& &vehicle routes & & & \\
\hline
Ferry replacement &Wireless &Nodes move &NI & $1^{(1)}$ &WSN \\
protocol \cite{1424832} &nodes &along routes & & & \\
\hline
IBRR \cite{1189163} &Satellites &Geosynchronous & &$>1$ & \\ \cline{1-3} \cline{5-5}

DSDV \cite{1247710} & & &FI &1 &Inter-planetary \\ \cline{1-1} \cline{5-5}
MRP \cite{1035720} & & & &  & Internet\\ \cline{1-1}
Combined DF &Wireless & Move among & & $>1$ & \\
scheme \cite{1139396} &nodes &groups or clusters & & & \\ \cline{1-1} \cline{4-6}
CAR \cite{1443499} & & &Both & $1^{(1)}$ & General \\
&&&&&communications \\
\hline
& &Between stations & & &Data backup and \\
& &according to track & & &connectivity to \\
TrainNet \cite{Araki2010} &Train &layout and &FI &1 &rural areas \\
& &timetable & & & \\
\hline
\multicolumn{6}{|l|}{(1) with load balancing, multiple nodes get to become a DF.} \\
\hline 
\end{tabular}
\end{center}
}
\end{sidewaystable}

\subsection{\label{STASEC}Statistical}
Proactive routing protocols rely on past or historical information such as temporal, spatial and nodes performance.  We will first outline works, see Section \ref{STA1}, that use temporal information before discussing those that exploit the spatial location and trajectory of nodes in Section \ref{STA2}.  Lastly, in Section \ref{STA3}, we present routing protocols that seek out nodes with good delivery ratios.

\subsubsection{\label{STA1}Temporal} 
In this category, a time-related function is used to aid forwarding decision. For example, the authors of \cite{4908299} use the encounter times of nodes.  This information is then used by nodes when forwarding bundles, where they preferentially select nodes with the smallest interval between rendezvous periods.   
In \cite{SushantJainAugust2004}, Jain et al. assume the availability of future contact periods.  They designed a metric called Minimum Expected Delay (MED) and modified Dijkstra algorithm to compute the path with minimal end-to-end delay.  As pointed out by Jones et al. \cite{Jones2007}, MED is only suitable for certain types of DTNs where future contact times are available, e.g., satellites.  To address this limitation, they proposed a new metric called Minimum Estimated Expected Delay (MEED), which is calculated using past contact history.  Each node then floods this metric throughout a DTN using a link-state routing protocol, which unfortunately, incurs a significant amount of overheads.  Moreover, it is unclear whether such an approach works in sparse DTNs.
Another example is Probabilistic Routing Protocol using History of Encounters and Transitivity (PROPHET) \cite{961272}, which uses a delivery predictability metric $P$. The metric has three features. Firstly, $P$ is computed iteratively using prior values.  Specifically, the $P$ value of node A to B is defined as,
\begin{equation}
P_(a,b)=P_{(a,b)_{old}}+(1-P_{(a,b)_{old}})\times P_{init}
\label{EQU5}
\end{equation}
where $P_{init}$ is the initial value, and $P_{(a,b)old}$ is the old value. Secondly, the value of $P$ degrades if nodes do not encounter each other within a given time interval.  In other words, 
\begin{equation}
P_{(a,b)}=P_{(a,b)_{old}}\times \gamma^\kappa
\label{EQU6}
\end{equation}
Here, $\gamma \in [0,1]$ is a fixed constant, and $K$ is the time units that have elapsed since the last update.  Lastly, the metric $P$ can be computed using those from other nodes. That is, if node A and B first meet each other, followed by node B meeting C, then 
\begin{equation}
P_{(a,c)}=P_{(a,c)_{old}}+(1-P_{(a,c)_{old}})\times P_{(a,b)}\times P_{(b,c)}\times \beta
\label{EQU7}
\end{equation}
where $0\le \beta \le 1$ is set manually to control the impact of transitivity.

In \cite{5474683}, the authors propose PROPHET$^+$, where a ``deliverability'' value, which is used to determine an appropriate routing path, is based on the following time-varying parameters: buffer size, power, bandwidth, and popularity.  Note that, popularity is defined as $1-\frac{N_t}{M_t}$, where $N_t$ and $M_t$ are the transmission rate and $M_t$ node's capacity in a given time interval respectively.    Their results show PROPHET$^+$ is able to reduce packet loss and delays because it only transmits bundles to nodes that have ample storage and power.  However, as pointed out by the authors, PROPHET$^+$ does not consider frequency of encounters. 

The authors of \cite{TC2010} showed via experimental studies that cumulative contact characteristics, e.g., probability of node contacts, do not capture transient contact patterns and connectivity.  For example, students at a university may meet for a prolonged period during class, and they may have a different set of friends when they are at home.  Any routing decision must therefore identify the correct subset of nodes when delivering bundles.  Moreover, when students are in class, they may form a connected network.  These students are said to have {\em indirect contacts}, and hence, present an ideal setting for bundle dissemination.  To this end, the authors propose a new metric that captures both direct and transient contacts within a fixed time interval. From trace-based simulation studies, they showed protocols achieved up to 50\% improvement in bundle delivery ratio.

In \cite{4558704}, the authors propose a protocol that is analogous to heat transfer.  Specifically, the delivery probability from nodes to sinks corresponds to the exchange of heat where nodes with a higher temperature transfer bundles to those with a lower temperature. The temperature of sinks is a constant $T$, which is set to a much higher value than other nodes. When nodes pass by a sink, they will be ``heated'', and hence, nodes with higher temperature has a higher probability of receiving bundles from other nodes.  The temperature of each node depends on their mobility and the frequency in which they visit sink nodes.  When nodes encounter each other, the one with the higher temperature will decrease in value, and vice-versa.  This means nodes with a lower temperature will send bundles to those with a higher temperature as they have recently visited a sink.  The authors deploy five nodes that move according to the random waypoint mobility model.  Their results show that nodes that are geographically closer to the sink have a higher delivery probability. However, the protocol has only been tested in a small network with five nodes. 

\subsubsection{\label{STA2}Spatial}
The nodes in this category record the speed, direction and mobility pattern of other nodes.    For example, Utility-based Distributed routing algorithm with Multi-copies (UDM) \cite{4625899} prioritizes nodes according to the number of connections a node has to their home communities. Here, ``home communities'' is defined as locations that nodes passed by and stayed close to most frequently.  This means these nodes are more likely to deliver bundles destined to nodes in a given home community. Apart from that, UDM uses binary transmission, where nodes send half of their bundles to another node as long as they have more than one bundle. Their experiments compare UDM with the epidemic and spray-and-wait routing protocol in terms of delivery rate and average delay.  Their results show that UDM decreases nodes' average delay by 500 seconds when transmitting 50 bundles, which is half that of spray-and-wait and a third of the average delivery rate achieved by epidemic routing protocol.  It is unclear whether similar results can be obtained for a different mobility model.  

Both \cite{1543743} and \cite{4656853} propose to exploit movement vectors. They assume that nodes have position awareness but do not know each other's movement pattern. The authors of \cite{1543743} propose the Motion Vector (MoVe) routing protocol, which uses node velocity and angle to calculate the shortest distance to a given destination.  Specifically, when they encounter each other, they compare their trajectory, and bundles are forwarded to nodes that are headed to the corresponding destination. Their results, from experiments comprising of 70 nodes with fixed source and destination nodes, show that increasing node numbers improves delivery rate and delay. Their experiments, however, do not consider the impact of duplicated bundles.  

Vector Routing (VeRo) \cite{4656853} uses the trajectory of nodes when forwarding bundles.   Specifically, nodes record their position and angle changes, and preferentially exchange a bundle with a node that is moving away from it.  
The main limitation of VeRo is that stationary nodes tend to receive the most bundles because every other nodes is effectively moving away from them.  
In a different work, the authors of Similarity Degree-based Mobility Pattern Aware Routing (SD-MPAR) \cite{5222878} assume that nodes with the same mobility patterns will tend to have similar movement angle and shorter distance between them; i.e., have stable relative positions. When two nodes encounter each other, they compare their similarity degree, which is a function related to the angle and distance between them. Nodes transmit bundles to those with a higher similarity degree. Their results show communication range to have an impact on delivery rate and average delay.  Moreover, their results show nodes that have a higher similarity leads to better delivery probability.  This, however, is at the expense of additional computational and storage capabilities.

The authors of \cite{1080146} exploit the mobility patterns of nodes when making routing decisions.  For example, a node will prefer to route bundles to nodes that are closer or headed toward the destination.  They propose to identify mobility patterns according to four functions.  That is, given the Cartesian coordinate of node $i$ and $j$, $(x_i, y_i)$ and $(x_j, y_j)$, they calculate their (i) Euclidean distance, (ii) Canberra distance, (iii) Cosine angle separation, and (iv) matching distance, where nodes are considered to have a similar distance if $x_j - x_i < \theta, y_j-y_i < \theta$, where $\theta$ is a predefined value. Their results show that the Euclidean distance and Cosine angle separation function have the best performance in terms of average delay.  As pointed out by the authors, the protocol can be improved further by incorporating temporal factors such as encounter duration or frequency. 

Lastly, Nelson et al. \cite{EC09} propose an Encounter-Based Routing (EBR) protocol.  Every node maintains a metric that reflects the average number of contacts within a given time interval.  This metric is then used to determine the number of bundle copies that is to be transmitted in each contact.  Specifically, a node with a high contact value will receive more copies of a bundle because it has a better chance of propagating a message, which leads to a higher bundle delivery ratio.  A key feature of the proposed routing protocol is that it bounds the maximum number of copies of a given bundle to $L$, where $L$ is either a fixed or probabilistic value.  This helps to keep resource consumption low, unlike works such as MaxProp \cite{4146881} and RAPID \cite{BALA2010} that require a high resource consumption in order to achieve comparable delivery ratio.

\subsubsection{\label{STA3}Stochastic}
The protocols in this category maintain a time varying network topology that is updated whenever nodes encounter each other.  For example, nodes using the Shortest Expected Path Routing (SEPR) \cite{1258396} protocol maintain a stochastic model of the network.  Each node constructs a time varying graph comprising of nodes they have encountered, and links that reflect the connection probability between nodes.  This also means nodes will exchange a link probability table containing past encounters whenever they meet other nodes.  Applying the Dijkstra algorithm on this graph, each node then calculates the expected path length to a given destination.  Their experiments demonstrate that SEPR has an improved delivery gain of 35\% with a 50\% reduction in resource consumption as compared to epidemic routing and its variants.  The key limitation of SEPR is that it has poor scalability due to its reliance on the Dijkstra algorithm.  Moreover, it is not suitable for DTNs with high node mobility.

MaxProp \cite{4146881} improves MV in the following manner. Firstly, MaxProp uses the hop count of bundles to better manage network resources. Secondly, acknowledgments are propagated through the network to erase expired bundles. Lastly, MaxProp deploys a mechanism to eliminate duplicated bundles. Similar to MV, MaxProp utilizes delivery probability during bundle transmission, in which each node initially defines their delivery probability as 
\begin{equation}
f=\frac{1}{|s|-1}
\label{EQU3}
\end{equation}
where $f$ is the initial delivery probability, $|s|$ is the number of nodes in the network. Upon encountering each other, nodes will update their delivery probability.  For example, given node A and B that exist in a network with $n$ nodes, where $f_A$ and $f_B$ represent their prior delivery rate, their new delivery rate $f'_A$ and $f'_B$ is calculated as
\begin{equation}
f'_A=f'_B=\frac{f_A+1}{\sum\limits^{n-2}_{n=1} f+f_A+1}
\label{EQU4}
\end{equation}
where $\sum\limits^{n-2}_{n=1} f$ represents the sum of other nodes' delivery rate.   Each node then maintains a sorted list of nodes delivery rates.  Bundles are then sent to nodes with the best delivery rate. Apart from that, MaxProp also employs the following prioritized bundle delivery schemes.  First, all bundles destined to neighbors are transferred first, followed by routing information, which includes the probability of meeting every other node.  After that, nodes deliver ACKs, followed by bundles that have not traversed far in the network.  Finally, nodes transmit the remaining packets and a hop-count list that reflects the list of nodes a bundle has already traversed.  MaxProp is evaluated using  a real-world testbed called UMassDieselNet.  The authors compared MaxProp to three routing algorithms.  Namely, (i) random transmission, (ii) Dijkstra algorithm, and (iii) most Encountered/Drop Least Encountered algorithm (ME/DLE) -- where bundles are transmitted to nodes that they encounter frequently, and are dropped if a bundle is destined for a node that the receiving node rarely meets.  
Their results show that bigger buffer size leads to an increased delivery rate. The key limitation of this protocol is that the initial delivery probability is a function of the number of nodes in the network.  Unfortunately, without global knowledge, this probability is inaccurate. 

Context-aware Adaptive Routing in Delay Tolerant Mobile Sensor Networks (SCAR) \cite{CARJ} calculates the delivery probability of a neighboring node according to the number of (i) encounters it has with sink nodes, (ii) connectivity with other nodes, and (iii) battery capacity.  In addition, SCAR is able to replicate a bundle to $R$ nodes with the highest delivery probability.  The authors, however, have not evaluated SCAR against any existing DTN routing protocols.

Balasubramanian et al. \cite{BALA2010} propose a routing protocol that considers network resources such as bandwidth and storage when optimizing a given route metric.   This is especially critical when nodes have resource constraints.  The said protocol, called Resource Allocation Protocol for Intentional DTN (RAPID), considers the utility of replicating a bundle at each rendezvous.   Here, the utility of a bundle models the benefits of replicating a bundle whilst taking into account resource constraints.  They also propose a control channel that allows nodes to collect information such as past transfers and encounters, which are then used to determine the utility $\frac{\delta U_i}{s_i}$ of a bundle $i$, where the numerator represents the increase in utility if the bundle is replicated, and $s_i$ is the bundle size.  Each node only forwards bundles with the highest utility amongst those in its buffer.  RAPID's performance is sensitive to how much and frequent information is distributed over the control channel.  Their trace-based studies show RAPID is able to deliver 88\% of packets with an average delay of 91 minutes.  These results, however, are achieved over one set of traces \cite{umass-diesel-2008-09-14}, and it is unclear whether similar performance can be achieved over other traces and indeed theoretical models.

\subsubsection{Social Networks}
Another approach taken by researchers to improve the performance of routing protocols is by identifying invariant properties in social networks.  Briefly, information diffusion in social networks is a well studied problem and bears resemblance to data transmissions.  As shown by Milgram in \cite{S.MilgramMay1967}, individuals tend to be separated by six degrees of separation.  In his famous experiment, 60 letters destined to a stockbroker were handed to a different person that only knows the stockbroker's name.  The study found that the median chain length of intermediate letter holders was approximately six.   This experiments demonstrates the ``small-world phenomenon'', and the ability to propagate information in a seemingly random network.

Henceforth, the authors of \cite{1143659} propose a routing protocol that exploits the fact that nodes in the real world routinely move and stay in several well known places.  This means nodes that visit these places have a contact probability and hence, will have more success delivering bundles.  Another example is \cite{4674358}, where Daly et al. propose three relevant properties of information flow: centrality, ties and predictors. \textit{Centrality} reflects a node's degree of importance, where a node with good centrality has a robust relationship with other nodes in the networks.  In particular, centrality can be further measured by three sub-degrees: Freeman's degree, closeness and betweenness. Freeman's degree is the number of directly connected nodes. Closeness refers to the path length or hop count between two nodes. Finally, betweenness corresponds to the number of times a node is used to relay bundles.   The \textit{ties} property evaluates the robustness of the connection between two nodes, where nodes with the following properties are deemed to have a high delivery rate: frequently connect at the same location, long encounter duration, and are well known to other nodes.  Lastly, the \textit{predictor} property indicates whether two nodes are likely to be connected to each other.  For example, as proven in \cite{Newman200539}, based on an analysis of the evolution of scientific collaborations, two co-authors of previous works tend to have a high probability of working together again in the future. In other words, previous connection is a good indicator of future ones.   In another work \cite{UserCentric2011}, Gao et al. consider user interests during routing.  They too exploit {\it centrality}, where they seek to exploit relays with high centrality in order to increase user satisfaction.  In \cite{Central2010}, the authors consider nodes with high centrality as well as their contact duration with others.  This is advantageous as popular nodes may only have brief encounters with a high number of nodes, and thereby, have limited communication capacity.  

\subsubsection{Discussion}
Table \ref{TAB5} summarizes the categories and metrics used by proactive routing protocols. To determine the next hop used for delivering bundles, protocols collect information such as time-varying metrics or geographical information.  Note that the latter information is easily obtained when nodes are equipped with a GPS unit. 

Works that exploit properties of social networks are summarized in Table \ref{TAB7}.  We can see all routing schemes utilize the notion of groups or communities where nodes are classified according to their common locations or hobbies.  Other than that, researchers also seek out popular nodes, as determined by their connectivity to other nodes.  These nodes, therefore, serve as a ``good'' next-hop when forwarding bundles.
%
%
\begin{table*} [htb]
{\scriptsize \caption{\label{TAB5}A summary of proactive routing protocols}} 
{\small
\begin{center}
\begin{tabular}{|l|l|l|} 
\hline
\multicolumn{1}{|c|}{\textbf{Protocol}} & \multicolumn{1}{c|} {\textbf{Category}} & \multicolumn{1}{c|}{\textbf{Metric}} \\
\hline
TIR \cite{4908299} & & Encounter times \\ \cline{1-1} \cline{3-3}
PROPHET \cite{961272} &Temporal & \\ \cline{1-1}
PROPHET$^+$ \cite{5474683} & &Time-varying delivery probability \\ \cline{1-1}
HEAT \cite{4558704} & & \\
\hline

UDM \cite{4625899} & & Location visit frequency \\ \cline{1-1} \cline{3-3}
EBR \cite{EC09}      & & Encounter frequency \\ \cline{1-1} \cline{3-3}
MoVe \cite{1543743} & Spatial & \\ \cline{1-1}
VeRo \cite{4656853} &  &Movement vector \\ \cline{1-1} \cline{3-3}
SD-MPAR \cite{5222878} & &Similarity of mobility pattern \\ \cline{1-1} \cline{3-3}
MobySpace \cite{1080146} & &Shortest distance to destination \\ 
\hline
SEPR \cite{1258396} & & Shortest path calculated by delivery  \\ 
                    & & probability \\ \cline{1-1} \cline{3-3}
SCAR \cite{1143656} & Stochastic & Connectivity change with other nodes \\
                    & & and remaining battery capacity \\ \cline{1-1} \cline{3-3}

RAPID \cite{ArunaBalasubramanian2007} & & Network bandwidth and node storage \\  \cline{1-1} \cline{3-3}

MaxProp \cite{4146881} & & Delivery probability  \\ 

\hline
\end{tabular}
\end{center}
}
\end{table*}

%
%
%
\begin{table*} [htb]
{\scriptsize \caption{\label{TAB7}A summary of social networks research}} 
{\small 
\begin{center}
\begin{tabular}{|l|l|l|l|} 
\hline
\multicolumn{1}{|c|} {\textbf{Studies}} & \multicolumn{1}{c|} {\textbf{Definition of}} & \multicolumn{1}{c|} {\textbf{Properties/Metrics}} & \multicolumn{1}{c|} {\textbf{Example}} \\
&{\textbf{a group}} & & \\
\hline
\cite{Pantazopoulos2010} & &Connections with & \\ \cline{1-1}
& &other nodes &  People that work in the same office \\
\cite{Hui2009} & Common & & or pass each other frequently \\
& location & &  \\ \cline{1-1} \cline{3-3}
\cite{Xu2009} & &Selfishness and &  \\
 & &altruism & \\
\hline
\cite{Boldrini2008} & Same hobbies &Social profile of & Club members \\
 & and locations &encountered nodes &  \\
\hline
& Social &Delivery probability & Family members or classmates or \\
\cite{Boldrini2008b}\cite{BBRAP} & relationship & and frequency to a &  popular individuals/communities \\
 &  & community &   \\
\hline
\cite{Boldrini2009a} & Locations & Visit frequency & People visiting the same shopping\\
 & & & mall \\
\hline
\cite{UserCentric2011} & Mobile users & User satisfcation & Individuals with high centrality; \\
                        & & & e.g., postmen \\
\hline
\end{tabular}
\end{center}
}
\end{table*}

\subsection{\label{CODER}Coding}
A standard approach taken by protocols we have reviewed thus far to improve bundle delivery ratio or their robustness is to increase the number of redundant transmissions.   The downside, however, is higher buffer occupancy level, which leads to packet loss.  Moreover, nodes may not be able to transmit all buffered bundles in each encounter.  To this end, a number of researchers have proposed to incorporate erasure coding, e.g., \cite{ECODE}\cite{LPDC}.  Briefly, given a bundle of size M and a replication factor r, an erasure coding algorithm produces M*r/b blocks of size $b$.   The key feature is that any $(1+\epsilon) \times \frac{M}{b}$ erasure coded blocks can be used to reconstruct the original bundle.   In other words, a receiver is able to recover a bundle after receiving a fraction of M*r/b coded bundles.   As a result, the dynamics of a DTN, such as uncertain contact periods and limited resources at nodes, have a lesser impact on bundle delivery.  As we showed later, this key characteristic increases the probability of bundle delivery whilst consuming less resources as compared to protocols that simply increase the number of duplicated bundles.

Wang et al. \cite{DTNCD1} proposed a novel forwarding method where coded-blocks are distributed widely to a number of nodes.  Advantageously, it limits each node to only a fraction of the coded blocks, which has the effect of constraining routing overheads.  The method works as follows.  Given a replication factor $r$, a source first divides a message into $k$ blocks, which are then coded into $k \times r$ coded blocks.  It then divides these blocks equally amongst $k \times r$ relays, where $k$ is a constant.  A receiver recovers a bundle completely after receiving $\frac{1}{r}$ of the coded blocks.  That is, a receiver re-constructs a message after meeting $k$ relays.  The authors showed via simulation and analytical studies that the use of erasure codes improves the worst case delay with a constant overhead.  In particular, the distribution of coded messages to a wide variety of relays help mask problems with a subset of these relays.   Moreover, unlike conventional routing protocols, the proposed method has constant overhead as it does not need to increase bundle redundancies to achieve comparable delivery ratios.

A fundamental problem in using erasure codes is determining the fraction of coded messages that should be transmitted amongst N paths such that the delivery probability is maximized.  This problem is made more difficult by the fact that each path has a different failure probability and throughput that are dependent on node capability and mobility.  Jain et al. \cite{DTNCD2} studied this problem over two cases.  The first is where the $N$ paths are governed by two models: Bernoulli and Gaussian.  The former model represents the scenario where a message is either fully lost or delivered, whereas in the latter, only a fraction of the messages are delivered.  To cope with the dynamic nature of DTN paths, the authors propose an optimization framework, and showed the Bernoulli case to be NP-hard.   Specifically, they propose a Mixed Integer Program (MIP) for the Bernoulli case, and adopt techniques from modern portfolio theory \cite{Port1} to attack the Gaussian case.   A limitation of this work is the use of a fixed-rate erasure code.  In this respect, the authors have not provided any methods to determine the optimal rate.  A key assumption in this work is that the availability of N paths that meet a given delay constraint, each of which has a well-known and constant probability over time.  In a different work, Tang et al. \cite{DTNCD9} addressed this limitation by considering a set of paths, each with varying stability.  They proposed to interleave coded blocks before transmitting them on available paths.  Their experimental results show sources using interleaving obtaining up to 30\% higher packet delivery ratio.  An interesting result is that delivery performance does not improve with increasing number of paths as coded blocks are more likely to be forwarded onto poor paths.   To address this limitation, the authors propose to send a different number of coded blocks on each available path as a future work.

In \cite{DTNCD11}, Liao et al.  used the contact frequency between a relay and destination node in order to decide the proportion of packets that are exchanged in each encounter.  In a subsequent work \cite{DTNCD12}, they extended their protocol to consider available resources at a given relay.  This is an important consideration as a relay with a high probability of meeting a destination may be low in power or buffer space.   The authors showed via simulation that the delivery ratio of the proposed protocols perform better than Spray-and-Wait \cite{1080143} and Wang et al. \cite{DTNCD1}'s method.   A caveat of these protocols, however, is that nodes need to learn the contact frequency of nodes.  Similarly, the authors of \cite{DTNCD10} employ erasure coding to help improve delivery with minimum overhead in a delay or fault tolerant mobile sensor network.    Specifically, their protocol determines the optimal erasure coding parameters such as the number of data blocks and redundancy given the current delivery probability.    

Wang et al. \cite{DTNCD2} and Jain et al. \cite{DTNCD2} have spurred a number of studies on using coding in DTNs.  In \cite{DTNCD3}, Altman et al. provided a closed-form solution that allows one to compute the delay and energy consumption of a DTN as a function of the coding used.  In this case, the authors considered erasure codes such as Reed-Solomon, and rate-less fountain codes.   The main limitation, however, is that they only consider two-hops routing schemes, where a source transmits a bundle to all encountered nodes, and relays only forward to destination nodes.  In \cite{DTNCD7}, the authors argue that coding based schemes are in-efficient in DTNs with good connectivity because of the additional information carried in coded blocks.   Moreover, as each relay only receives a limited number of coded blocks, existing approaches waste the remaining time of a given contact.   To this end, they propose two protocols: (i) Aggressive Erasure Code (A-EC), where a node transmits as many coded blocks as possible in every contact duration, and (ii) Hybrid Erasure Code (H-EC), where a node transmits two copies of a coded block.   In other words, the first copy is sent as per \cite{DTNCD1}, and the second copy is transmitted using A-EC.  A disadvantage of both protocols is their higher traffic overhead.  In a different work, Vellambi et al. \cite{DTNCD8} use rateless codes to improve delivery reliability.  In particular, they aim to overcome the negative impact of discarding expired bundles, a standard approach used to reduce buffer occupancy level.  Unlike \cite{DTNCD2}, which used a fixed-rate erasure code, the authors consider rate-less or fountain codes; see \cite{LTCODE}.  Their simulation studies over Random Waypoint Model and UMassDieselNet dataset [ref] show better performance over replication based protocols, and the method proposed by Wang et al. \cite{DTNCD1}.  An interesting finding is that reliability and latency degrade gracefully with decreasing expiry time.

Bulut et al. \cite{DTNCD13} study the cost of erasure coding, where cost is defined as the total bytes transmitted between a source and destination pair.  In particular, they study the cost incurred by the following spraying algorithms:  binary and source \cite{1080143}.  They found messages are distributed more slowly when nodes use source as compared to binary spraying.  However, binary spraying incurs high overheads.  To this end, the authors designed a protocol that adjusts the replication factor r and number of blocks k, see \cite{DTNCD1}, dynamically based on two factors: (i) a node's probability of meeting a destination node within a given deadline, and (ii) the required delivery ratio.  In addition, they consider spraying messages at different periods, each with its own replication factor that is a function of the number of messages that has arrived at the destination in the last period.   Their method, however, requires nodes' contact history and relies on acknowledgments from a destination node, which may not be forthcoming.

In \cite{DTNCD1}, Widmer et al. studied DTNs where no oracles are available to provide topological information.  In this type of DTNs, conventional approaches tend to employ probabilistic routing protocols; see Section \ref{STASEC}.  The authors improve upon these protocols using network coding \cite{ NBase}.   Consider the following simple example comprising of nodes placed in a linear topology:  A-B-C.   Assume node A transmits packet {\bf a} to C, and conversely, node C sends packet {\bf c} to node A.   When node B gets both packets, it broadcasts packet {\bf b}, where $b = a \oplus c$.  Node A can then retrieve packet {\bf c} as follows: $c = b \oplus a$.  A similar operation is carried out by node C to retrieve packet {\bf a}. It is worthwhile pointing out that unlike conventional mobile ad-hoc networks, e.g., \cite{KATA1}, the probability of having multiple nodes within the transmission range of a node is exceedingly small in DTNs.  Widmer et al. adopt linear network coding \cite{NBase2}, where a node sends a linear combination $\sum_i g_i x_i$, where $x_i$ is the packet of interest.  Here $x_i$ and $g_i$ are interpreted over a finite field.  A nice property of linear network coding is that n packets can be decoded if a node receives $m$ combinations, and $m$ has rank $n$.   Effectively, these combinations can be viewed as independent equations which form a system of linear equations that can then be inverted to reveal the n packets.  To this end, the authors designed a scheme that ensures a sufficient number of combinations are propagated by each node such that packets can be decoded with a high probability whilst keeping memory requirement at a minimum.   More specifically, for a given forwarding factor $d$, a node generates $\lfloor d\rfloor$ information vectors, and an additional vector is generated with probability $d$ - $\lfloor d\rfloor$.  From extensive simulation results, the authors showed that the proposed scheme performs better than probabilistic algorithms in terms of packet delivery in extreme conditions.  Moreover, they observed that the proposed algorithm benefited from node mobility.  A key future work is extending their scheme to consider the case where non sink nodes are not required to decode data.

In \cite{DTNCD4}, the authors study the impact of Random Linear Coding (RLC) on the delivery delay of epidemic routing protocols subjected to bandwidth and buffer constraints.   At each encounter, nodes exchange their encoding matrix to determine whether any combinations will lead to an improvement in rank.  If so, these combinations and random coefficients are then exchanged.  Moreover, an anti-packet is generated once a node decodes all packets.  The authors studied two cases: (i) a single block of packets destined to one or more receivers, and (ii) multiple coded blocks.  They showed that RLC improves the delivery and delay performance of epidemic routing protocols in both bandwidth and buffer constrained cases - a similar conclusion was also put forth by Lin et al. \cite{DTNCD14}.  In \cite{DTNCD4}, the authors showed via an analytical framework that network coding yields a significant advantage in terms of reducing delay.  In addition, they proposed a priority coding protocol that classifies K packets into different priority levels.  Each level is then coded and transmitted.  Once an ACK is received by the source, it codes and transmits packets in the next priority level.

Ahmed et al. \cite{DTNCD6} exploit the power law behavior observed in DTNs comprising of humans carrying personal communication devices \cite{POCKET}.  Another example is vehicle based DTNs \cite{4146881}.   This means a small number of nodes have a high degree of connectivity.   In other words, they make good relays or hubs as all other nodes are reachable via them over a small number of hops.  To this end, Ahmed et al. designed a protocol whereby nodes forward bundles to hubs, which then construct a linear combination of these messages to reveal an encoded message.  This is carried out to take into account the bounded bandwidth of each hub.  Hubs exchange coded messages that increase their matrix rank whenever they meet.  The resulting protocol is shown to have 20\% higher delivery ratio as compared to Epidemic \cite{Vahdat2002}, Spray-and-Wait \cite{1080143}, RLC \cite{DTNCD4}, and BubbleRap \cite{BBRAP} in a DTN comprising of 1163 buses operating in Seattle \cite{BUS2003}.

\subsubsection{Discussion}
Existing works can be divided into source and network coding based approaches.  The main problem with source based approaches is that intermediate or relay nodes may waste bandwidth transmitting duplicated coded packets.   A key disadvantage of both approaches is decoding delay, which thus far, only Lin et al. \cite{DTNCD5} have addressed.   Prior works can also be distinguished by their assumptions on DTN dynamics.  That is, whether they consider prior knowledge in terms of contact duration and times between nodes, and whether a source is aware of the delivery probability at a destination.  Also, coding approaches have primarily been evaluated in the context of epidemic based routing protocols.  Therefore, it will be interesting to investigate their performance in DTNs that use protocols outlined in Section \ref{DFSEC} and \ref{STASEC}.  Indeed, a key future work will be to evaluate their performance in conjunction with mechanisms such as the use of TTL, EC and anti-packets.  Lastly, apart from \cite{DTNCD13} and \cite{DTNCD11}, little work has considered adjusting the parameters $r$ and $k$ dynamically in accordance with the dynamic nature of a DTN. 

\section{\label{MSEC}Multicast}
Applications may need to deliver bundles to a group of users.  For example, the dissemination of software patches and targeted advertisements.   Supporting multicast in DTNs is non trivial as there are frequent link partitions and nodes experiencing unpredictable transmission delays.  This means it is unlikely that nodes or subscribers will receive bundles at the same time, nor for source(s) to receive all acknowledgments.  In addition, group membership changes, varying node speeds and density further add to the complexity of designing multicast protocols that run well in DTNs \cite{5205611}.
Apart from that, multicast poses a fundamental problem to statistical protocols, see Section \ref{STASEC}, as existing methods select relays based on their delivery probability to {\em one} destination only.
Multicast protocols for DTNs can be classified into two categories: (i) unicast-dependent, and (ii) statistical.

\subsection{Unicast Dependent}
Zhao et al. \cite{1080145} define three kinds of multicast receivers. The first is called Temporal Membership (TM), where nodes that are connected in a given time period are viewed as members belonging to the same group.  The second is called Temporal Delivery Model (TD), which combines TM with a delivery threshold for group members.  This means, within a given time period, nodes are considered group members if they have connectivity to each other and can transmit bundles to their destination within a delay threshold.  The last one, called Current-Member Delivery Model (CMD), considers the stability of group members -- i.e., nodes that remain static for the duration of a multicast session.   
Zhao et al. also proposed four fundamental multicast routing algorithms: 
\begin{enumerate}
\item Unicast-Based Routing (UBR) -- where unicast is used to emulate multicast whereby a source node sends bundles via unicast to each group member.  
\item Static Tree-Based Routing (STBR) -- a spanning tree is constructed by a source node, and the tree remains static for the duration of the multicast session. This means if the path connecting node A and B breaks, these nodes will not be able to communicate. 
\item Dynamic Tree-Based Routing (DTBR) -- where each node and bundle are allocated a group ID, and bundles are only received by group members with the same group ID. DTBR chooses the shortest routing path to each group members using the Dijkstra algorithm.  
\item Group-Based Routing (GBR) - where nodes with contemporaneous paths are designated as a group, and a source multicast bundles to each group using STBR. Within a group, flooding is used to deliver bundles.  
\end{enumerate}

All four categories of multicast routing protocols have several shortcomings. UBR requires ferries to meet every subscriber.  This means UBR does not adequately utilize the mobility of all non-ferry nodes to help improve the delivery of bundles. Both STBR and GBR have poor flexibility. In STBR, all routing paths are static, and if the path connecting node A and B breaks, both nodes will not be able to communicate. In GBR, bundles are delivered to a group member, which then floods the bundles to other group members. However, GBR still uses STBR to deliver bundles.  As DTBR uses Dijkstra's algorithm, it requires a DTN to have contemporaneous paths between nodes. 

The authors of \cite{Ye2009} investigate a variant of DTBR called On-demand Situation-aware Multicast (OS-Multicast).  One significant difference is that nodes build a dynamic multicast tree according to their history of encounters. That is, nodes maintain encounter records and duration with other nodes, and use this information to calculate the shortest path to a node. The authors evaluate the performance of OS-Multicast, UBR and STBR in terms of message delivery rate, delivery efficiency and average delay. Their results show that OS-Multicast has better delivery rate and less delay. Moreover, OS-Multicast has a higher reliability as compared to DTBR as there are more duplicated bundles in the network. 

The authors of Context Aware Multicast Routing (CAMR) \cite{1163622} propose a multicast routing scheme that is based on DTBR but with the addition of a route recovery mechanism. Nodes multicast bundles through a recently computed shortest path -- as per the Dijkstra algorithm.   Apart from that, nodes broadcast neighbor discovery messages and if the number of neighbors fall below a threshold, nodes set the ``sparsely connected'' flag and increase their transmission power to recruit more neighbors.  In terms of delivery rate, CAMR delivers 1.3\% more bundles as compared to DTBR.  In addition, the authors reported a 23\% decrease in average delay.  However, the limitation of CAMR is that the threshold value is fixed and not adaptable to changes in node movements. 

A critical problem for all the aforementioned multicast schemes is the reliance on a spanning tree. However, in a network where topologies change frequently, nodes experience long disconnection times, and may have no acknowledgments from other nodes.  Hence, generating and maintaining a spanning tree in such a network becomes impossible. In epidemic routing protocol and its variants, bundles transmissions are based only on node encounters, and does not rely on a spanning tree. Moreover, bundles are propagated at a higher rate, and hence nodes experience a better delivery ratio as all nodes participate in the forwarding process.  This is in contrast to data-ferries based multicast routing schemes, where bundles are delivered by designated nodes.

\subsection{Statistical}
In \cite{4770551}, Abdulla et al. argue that the aforementioned approaches are not readily applicable in DTNs due to the lack of knowledge regarding node connectivity and mobility. In other words, it is difficult to form a multicast tree that adapts to the vagaries of DTNs.  To this end, they propose Controlled Epidemic Routing for Multicast (CERM). Source nodes multicast bundles to all encountered nodes until they are received by all subscribers. The key problem is duplicated bundles.  In this respect, the authors propose two ways to eliminate these bundles. First, they propose the use of synchronization servers, which keep track of TTL value of bundles.  Once a bundle expires, these servers flood control messages to all nodes to purge the bundle.  The second proposal involves embedding a TTL value in each bundle.  Their results show that delivery rate is proportional to multicast group size and simulation time. The group size, however, has no impact on the average delay.  This work, however, assumes all nodes have synchronized clocks.  In addition, it is unclear how TTL can be adapted to the varying dynamics of a DTN.

The fundamental problem addressed by Gao et al. \cite{MCAST09} is to select the minimum number of relays that ensures a given bundle delivery ratio to multicast receivers.  Interestingly, this is effectively a variant of the well known knapsack problem, where each item $k$ or relay has a weight $w_k$.  To calculate $w_k$, the authors employ two key social characteristics, centrality and communities.  That is, popular relays or those that are part of the same community as multicast subscribers have a higher weight.  Using trace-based simulation, the proposed protocols, Single-Data Multicast (SDM) and Multiple-Data Multicast (MDM), are shown to have 20\% and 50\% higher bundle delivery ratio respectively than pure epidemic \cite{Vahdat2002} and PROPHET \cite{961272}, and requires fewer relay nodes.

In the DTN Pub/Sub Protocol (DPSP) \cite{4483167}, there are two types of node: publisher and subscriber.  Nodes designated as publisher store and classify bundles, and flood them to subscribers.  Each node maintains a subscription list to record the multicast group they belong to, and floods this list to all other nodes. If nodes receive a bundle destined to a group they have not subscribed to, they will store and forward the bundle to nodes that are subscribed to the corresponding group.  Their results show that DPSP has a better delivery rate but worse delay than epidemic routing.  The limitation of DPSP is that nodes use flooding to announce their subscription.   Hence, it is not scalable with increasing node numbers. 

In order to support multicast, the authors of \cite{1683552} propose to use either epidemic routing protocol or DFs depending on group numbers.  Source nodes first send bundles to be multicast to DFs, which then decide, according to group size, whether to deliver these bundles one by one or to use the epidemic routing protocol.  In terms of delivery rate and delay, their experiments show the proposed protocol reaches a delivery rate of 75\%.  However, they also showed that the average delay decreases when the group size increases.   

\subsection{Discussion}
Table \ref{TAB6} summarizes the mechanisms and metrics used by each of the aforementioned multicast routing protocols. Additionally, we indicate their delivery rate and delay.   Thus far, these protocols assume the ability to build a spanning tree.  Hence, there are only applicable to DTNs with semi-static nodes.  Moreover, little to no work has investigated the advantages and disadvantages of using epidemic and DFs routing protocols.  We elaborate on these observations further in Section \ref{EIDEA3}.

\begin{sidewaystable}[htbp]
{\scriptsize \caption{\label{TAB6}A summary of multicast routing protocols}} 
{\small 
\begin{center}
\begin{tabular}{|l|c|c|c|c|} 
\hline
\multicolumn{1}{|c|} {\textbf{Protocols}} & \multicolumn{1}{c|} {\textbf{Multicast Mechanism}} & \multicolumn{1}{c|} {\textbf{Metrics}} & \multicolumn{1}{c|} {\textbf{Limitations}} & \multicolumn{1}{c|} {\textbf{Group}} \\
&\multicolumn{1}{c|} {\textbf{}} & &\multicolumn{1}{c|} {\textbf{}} &\multicolumn{1}{c|} {\textbf{Definition}} \\
\hline
\textbf{UBR \cite{1080145}} &Calculates a separate routing path  &As per the cost used  &Consume  significant resources  & \\ 
 &from a source to each group members &by unicast protocol &if multicast group is large & Nodes which  \\ \cline{1-4}
 
\textbf{STBR \cite{1080145}} & Constructs a static spanning tree  & Bandwidth  & Static spanning tree is not   & require  \\ 
                             &  & or transmission rate & suited for dynamic DTNs &the same bundles \\ \hline

\textbf{DTBR \cite{1080145}} & Uses Dijkstra algorithm  & Transmission cost  & Bundles cannot be  & Same group ID \\ 
														&  & between nodes  & transmitted between groups & \\
														&  & in the same group & & \\ \hline

\textbf{GBR \cite{1080145}} & Uses STBR between groups and   & Transmission cost  & Uses flooding within group,  & \\ 
														& flooding within groups  &between groups &which consumes  & \\ 
														&  & &	significant amount of resources & \\ \cline{1-4}

\textbf{OS-Multicast }& Uses DTBR with historical  &As per node's  & High signaling overheads  &Nodes which  \\ 
\textbf{\cite{Ye2009}}& encounters  & destination list &associated with  & require the \\ 
																	  & & &multicast tree maintenance & same bundles \\ \cline{1-4}

\textbf{CERM \cite{4770551}} & Employs epidemic routing, and is able  & Expiration time & Difficult to synchronize   & \\ 
														 &  to eliminate duplicated bundles & & all nodes in DTNs  & \\ 
														 & & &due to frequent disconnections & \\\hline

\textbf{DPSP \cite{4483167}} & Relies on flooding and group   &Subscription list  &Uses flooding to announce  &Nodes with the same   \\ 
														 & member filtering & &nodes' subscription & subscription that describes  \\
														 & & & & required bundles' categories   \\  \hline

\textbf{CAMR \cite{1163622}} & Uses DTBR with a recovery  & Neighbor counts &Does not work well in sparse  &Nodes in  \\
														 & mechanism & & networks due to the cost of  & the same location \\ 
														 & & & neighbor search & \\ \hline

\textbf{MEPDF \cite{1683552}} & Combines epidemic routing and DFs,  &Group size &Static threshold value  &Nodes that require  \\
															& and adaptively chooses suitable   & &for group size &the same bundles \\
															& protocols according to group size & & & \\ \hline

\textbf{SDM, MDM \cite{MCAST09}} &  Forward bundle to relays with  & Cummulative  &  Relays must be in contact            &  Nodes that require \\ 
                                 &  highest centrality and part of & probability to             &  with source by the given & the same bundle \\ 
                                 &  a community with multicast     & multicast subscribers            &  deadline                                         & \\ 
                                 &  subscribers.                   &             &                                                   & \\\hline

\end{tabular}
\end{center}
}
\end{sidewaystable}

\section{\label{ESEC}Future Research Areas}
%
%
%
In this section, we outline efforts that seek to adapt results from social networks and epidemiology in the hope of increasing the delivery ratio and minimize delay of bundles in DTNs that use humans as a method of transport.  Then, in Section \ref{EIDEA2}, we briefly discuss the lack of uniform research methodologies that allow researchers to compare all routing protocols objectively.  After that in Section \ref{EIDEA3}, we discuss potential research directions of multicast protocols.
\subsection{Epidemiology}
The field of epidemiology studies the spread of disease in scale-free or social-contact networks.  In this respect, a promising research direction is to exploit similar concepts and models.  As Table \ref{TAB8} shows, DTNs and epidemiology share key concepts that we can exploit to create data distribution protocols.  By mining social networks for contact patterns, and learning their dynamic properties, we can formulate design guidelines that would lead to practical and efficient data dissemination protocols.  Watts and Strogatz \cite{REF28} studied the identification of individuals with the highest ``centrality'' -- a measure of contacts, the proportion of time an individual lies on the shortest path between other individuals, and the average steps between an individual and all other individuals.   This means nodes can monitor the centrality of encounter objects in the hope of identifying those that maximize delivery ratio as well as minimize delay.  Another interesting work which researchers can exploit is a study by Brockmann et al.'s \cite{REF32}, which uses a Web game, called \textit{Where's George?} (http://www.wheresgeorge.com/), to track the geographical location of dollar bills, and from that derives statistical laws of human travel.  Here, ``dollar bills'' are analogous to nodes carrying messages.  Hence, by adapting the models in \cite{REF32}, researchers will be able to analyze the following properties:  the frequency and duration of contact periods between nodes, packet delivery ratio, and message delivery delays.  Another potentially relevant work is that of Eubank et al. \cite{REF29}, where they model contact networks using bipartite graphs.  Their model describes people's mobility and locations, whereby two persons in the same location are likely to be infected.  Eubank et al. then use the resulting model to analyze the spread of diseases and also to develop methods for surveillance and vaccination.  This work shows that location can be used to cluster nodes that are likely to pass a given geographical area.  

\subsection{\label{EIDEA2}Theoretical and Trace Studies}
An important observation is that the works we have reviewed thus far, most if not all, use the random way point mobility model.  
Table \ref{TAB9} lists the experiment parameters of all protocols.  
There is thus a need to experiment with more realistic models.  This issue is particular critical given that DTNs are usually based on the movements of both inanimate and animate objects; e.g., vehicles and animals.
Apart from that, we can see all experiments contain a small number of nodes with limited data rate in large areas for simulating sparse nodes density. Additionally, nodes are also allocated limited transmission range and buffer size.   Given that all existing works use varying simulation methodologies, it is therefore very difficult to determine the ``best'' performing routing protocol for a given DTN.  Therefore, there is a critical need for a unified research methodology that compares all routing protocols comprehensively.

To date, only a handful of works have proposed alternative mobility models.  Bai et al. \cite{1208920} introduce several mobility models; namely, group, freeway and Manhattan model.  The group model is also called the Reference Point Group Mobility Model (RPGM) \cite{313248}. In RPGM, nodes move together as a group and every group has a central node called group leader. The group leader determines the movement of group members.  In the freeway model, nodes move according to predetermined routes.  Lastly, the Manhattan model simulates nodes movement in a metropolitan scenario.  Other than that, Leung et al. \cite{329340} describe a highway model where nodes enter and leave a highway through multiple entrances and exits.  Lastly, the Homing-Pigeon-Based (HoP) model \cite{5282549} models a scenario where each community has a designated message deliverer that periodically carries bundles from their home community to various destinations before returning home.  These works, however, have not comprehensively compared routing protocols designed for DTNs, and thus is an important future work.

Recently, researchers have begun using trace based simulation studies.  That is, instead of using theoretical models such as Random Way Point, they use traces of node movements at a given location; e.g., at a conference or city.  It is important to note that trace files are specific to a given environment, and cannot be readily generalized to other scenarios.  Moreover, they have limited number of nodes; i.e., they cannot be scaled readily to thousands of nodes.  Their advantages, however, include the availability of contextual information, and group or community membership.  Also, based on nodes' pre-recorded movements, one can easily determine nodes with high {\it centrality}, and the optimal forwarding path.  Indeed, this is the key observation that motivated Hui et al. \cite{BBRAP} to develop a forwarding algorithm, called BUBBLE.  They identified via trace-files analysis that people based DTNs are characeterizedby popular individuals or groups.  In effect, individuals and communities have a ranking that denotes their 'popularity' in a given DTN.  To this end, their algorithm fowards or ``bubbles' bundles to increasingly popular nodes or communities.

Past works usually use traces that capture a node's location, encounter times and durations.  For instance, in \cite{epfl-mobility-2009-02-24}, researchers traced the positions of 500 taxis in San Francisco over 30 days using GPS and roadside servers with wireless transmitter.  In a different work, the authors of \cite{mit-reality-2005-07-01} collected the ID of Bluetooth devices carried by students on a university campus. By far, the following are the most popular trace files: (i) Dartmouth/campus \cite{dartmouth-campus-2009-09-09}, (ii) Haggle \cite{cambridge-haggle-2009-05-29}, (iii) MIT/reality \cite{mit-reality-2005-07-01}, (iv) National University of Singapore (NUS) \cite{nus-contact-2006-08-01}, and (v) UMass/diesel \cite{umass-diesel-2008-09-14}.   
However, thus far, there has been a lack of work that compare the performance of DTN routing protocols using a variety of traces.  Lastly, a critical issue is that no works have used a mixed of theoretical and trace based simulation studies to evaluate proposed protocols.

\begin{sidewaystable}[htbp]
{\scriptsize \caption{\label{TAB8}Analogous concepts in epidemiology and the challenged networks}} 
{\small
\begin{center}
\begin{tabular}{|l|l|l|} 
\hline
{\textbf{Concept}} & {\textbf{Epidemiology}} & {\textbf{DTNs}} \\ \hline
{\textit{Information Carrier}} & Viruses & Nodes \\ \hline

{\textit{Performance Metrics}} & Spreading rate             & Number of bundles delivered \\ 
                               & Number of people infected  & Number of devices with bundles \\ \hline

{\textit{Application Constraints}} & Virus lifecycles             & Bundle delivery deadline \\ \hline

{\textit{Host Constraints}}     & Opportunity for infection and contagiousness   & Node's transmission range and data rate. \\ 
                                & A person's medical profile; e.g., age, gender. & Node's storage capacity \\
                                & An individual's susceptibility to infection    & Node already has stored bundle \\ \hline
                                
{\textit{Congestion Control}} & Vaccination methods               & Removing bundles from nodes \\ \hline

{\textit{Topology}} & Population's spatial distribution           & Node's spatio-temporal patterns \\ \hline

\end{tabular}
\end{center}

}
\end{sidewaystable}

\begin{table*} [htb]
{\scriptsize \caption{\label{TAB9}A summary of experiment parameters}} 
{\small 
\begin{center}
\begin{tabular}{|l|l|l|l|} 
\hline
\multicolumn{1}{|c|}{\textbf{Protocols}} & \multicolumn{1}{c|}{\textbf{Epidemic routing}} & \multicolumn{1}{c|}{\textbf{DFs}} &\multicolumn{1}{c|}{\textbf{Proactive}} \\
\hline
\multicolumn{1}{|l|}{\textbf{Number of Nodes}} & $\le$100 & $\le$ 120 & $\le$ 240 \\
\hline
\multicolumn{1}{|l|}{\textbf{Mobility pattern}} &Random waypoint & Random waypoint, &Random waypoint \\
& &along streets & \\
\hline
\multicolumn{1}{|l|}{\textbf{Area}} & $\le$ 50km$^2$ & $\le$100km$^2$ & $\le$25km$^2$ \\
\hline
\multicolumn{1}{|l|}{\textbf{Data rate}} & \multicolumn{2}{c|}{$\le$10KBps} & $\le$250KBps \\
\hline
\multicolumn{1}{|l|}{\textbf{Transmission}} & \multicolumn{3}{c|}{ $\le$300m} \\
\hline
& &Delivery rate, & \\
&Delivery rate, &average delay, & \\
\multicolumn{1}{|l|}{\textbf{Evaluation}} &average delay, time &KB/J, number of &Delivery rate, \\
\multicolumn{1}{|l|}{\textbf{metrics}} &to deliver all &times DFs move &average delay \\
&bundle &from sources to & \\
& &destination nodes & \\
\hline
\multicolumn{1}{|l|}{\textbf{Buffer size}} & \multicolumn{3}{c|}{Infinite or up to 5MB} \\
\hline
\multicolumn{1}{|l|}{\textbf{Bundle size}} &\multicolumn{3}{c|}{$\le$14MB} \\
\hline
\end{tabular}
\end{center}
}
\end{table*}

\subsection{\label{EIDEA3}Multicast}
To date, there are only a few multicast routing protocols.  Hence, there are ample opportunities to design new ones that incorporate the various strategies used by epidemic, DFs and proactive based routing protocols.
In particular, only a few works have studied the efficacy of epidemic routing protocols in delivering bundles to multicast group members. Thus far, in \cite{4770551}, the authors have only investigated epidemic with TTL. The authors, however, have not considered multicast sessions with a large number of subscribers, where a small TTL value may lead to bundles expiring prematurely.  This means epidemic with EC or epidemic with immunity are better suited for multicasting in DTNs. However, currently, no work has conducted any investigation on these two buffer policies.  

Second, no researchers have studied the effect of anti-entropy in multicast scenarios.  In \cite{Vahdat2002}, the authors pointed out that `comparing before exchanging' is critical because it avoids duplicated bundles.   This is important because once a node receives a bundle, it will never receive the bundle again even though said bundle has been discarded from its buffer. However, in multicast, given that there may be more than one subscriber, it is unclear whether anti-entropy will result in a lower bundle delivery ratio.  Intuitively, a node that has delivered a bundle, and subsequently deleted the bundle before meeting another subscriber should be given another copy of said bundle to improve bundle delivery ratio. 

Third, no research has analyzed the impact of having multicast subscribers act as relay nodes to forward bundles. In every epidemic variant, e.g., \cite{4770551}, \cite{962117}, \cite{4557809}, subscribers are effectively sink nodes, and are not required to forward bundles to others. However, in multicast, it is unclear how this influences multicast delivery.   This is an important consideration when there are more subscribers than non-subscribers or relay nodes.  If there are only a few relay nodes, the probability of a subscriber meeting a relay node becomes less, which increases end-to-end delays.  In the worst case, only the source is the relay and all other nodes are subscribers.  This means the source will have to meet each subscriber in order to deliver bundles.

Fourth, the size of a multicast group affects the delivery ratio of epidemic with TTL \cite{4770551}. Their experiments show that when the multicast group size increases, the delivery ratio drops accordingly.   In particular, when the multicast group size increases from two to 60, delivery ratio reduces from 95\% to 50\%.   However, it is unclear whether multicast group size has any impact on delivery ratio if subscribers do not forward bundles, or when nodes employ EC or immunity packets to discard bundles.  This is because the ratio of multicast subscribers to non-subscribers is a critical issue as larger multicast group sizes mean fewer relay nodes, and vice-versa. Moreover, in epidemic with immunity, the number of immunity packets corresponds to the multicast group size. Increasing the multicast group size translates to proportionally more immunity packets. This, however, corresponds to faster bundle deletion despite some subscribers not having received the deleted bundles. 
\section{\label{CONC}Conclusion}
The concept of DTNs has matured significantly in the last decade, with a myriad of novel applications that range from tracking wildlife to exploiting the movement of humans.  A notable application is providing connectivity to rural areas that otherwise would have been isolated from the Internet.  As a result, like the Internet, it is almost impossible to predict what economical, social and technological impact DTNs will have in the future.

A key underpinning that will ensure the success of DTNs is the performance of routing protocols.  In this respect, this paper has reviewed three main categories of routing protocols, where we provide a comprehensive qualitative comparison of key features within as well as between categories.  Furthermore, we presented the handful of works that aim to design efficient multicast routing protocols.  We also highlighted the research methodologies used by existing works and the lack of a common set of parameters that can be used to evaluate existing protocols objectively.  Finally, we review an emerging area that aims to incorporate research from social networks and epidemiology in the hope of identifying invariant properties that will help enhance the performance of routing protocols.  We believe this to be the next frontier in DTNs research, especially in light of ubiquitous, powerful mobile devices that are equipped with ample storage, camera and positioning capability that are capable of carrying data as well as collecting context and social information.

\bibliographystyle{abbrv}

\end{document}